\documentclass[11pt]{amsart}
\usepackage{latexsym,amssymb,amsmath,amscd,amsthm}
\numberwithin{equation}{section}
\theoremstyle{remark}

\newtheorem{proposition}{{\bf PROPOSITION}}[section]
\newcommand{\bq}{\begin{equation}}
\newcommand{\bea}{\begin{array}}
\newcommand{\eea}{\end{array}}

\newcommand{\ga}{\alpha}
\newcommand{\gep}{\epsilon}
\newcommand{\gD}{\Delta}
\newcommand{\gl}{\lambda}
\newcommand{\gL}{\Lambda}
\newcommand{\gb}{\beta}

\newcommand{\mf}{\mathfrak}
\newcommand{\mc}{\mathcal}

\newcommand{\ul}[1]{\underline{#1}}

\newcommand{\go}{\omega}
\newcommand{\gO}{\Omega}
\newcommand{\gG}{\Gamma}
\newcommand{\gt}{\theta}
\newcommand{\gs}{\sigma}

\newcommand{\gag}{\gamma}
\newcommand{\gd}{\delta}
\newcommand{\pp}{\partial}

\newcommand{\tl}{\tilde}
\newcommand{\na}{\nabla}
\newcommand{\gk}{\kappa}

\newcommand{\bs}{\blacksquare}

\newcommand{\gS}{\Sigma}

\newcommand{{\DDD}}{D\!\!\!\!\!\!-}

\newcommand{\bx}{\Box}

\title{GRAVITY AND THE QUANTUM POTENTIAL}
\author{Robert Carroll\\University of Illinois, Urbana, IL 61801}

\date{May, 2004\thanks{email: rcarroll@math.uiuc.edu}}

\begin{document}

\bibliographystyle{plain}

\begin{abstract}
We review some material connecting gravity and the quantum potential and provide a few new
observations.  The main theme is that in utilizing a conformal (Weyl) geometry the metric plus the
Weyl vector plus the quantum mass field determine spacetime geometry.  There are strong connections
to deBroglie-Bohm style ideas in quantum theory.
\end{abstract}

\maketitle

\tableofcontents

\section{INTRODUCTION}
\renewcommand{\theequation}{1.\arabic{equation}}
\setcounter{equation}{0}

We sketch here first some results extracted from
\cite{b98,s31,s32,s33,s34,s35,s36} on relativistic Bohmian mechanics,
Weyl geometry, and quantum gravity (cf. also
\cite{a21,b40,b3,b37,b38,b70,b8,b7,ch,c2,c3,c4,c5,c6,c7,c8,c9,d4,d42,d15,f2,f3,f7,f8,f9,g50,
h99,h98,m7,m1,m2,s2,s4,s6,s7,ss3,ss4,ss5,s37,s38}).
We use the Bohm-Weyl terminology to refer to a series of papers by Albohasani, Bisabr,
Darabi, Golshani, Motavali, Salehi, Sepangi, A. Shojai, and F. Shojai dealing with the subject; it
could perhaps be called the Tehran approach or named after some group of authors.  However
the idea of linking Weyl geometry and deBroglie-Bohm type quantum theory also does appear elsewhere
as indicated in this paper and was perhaps first noticed in \cite{lo} (cf. also \cite{or}).  
The "Tehran" version is consolidated and summarized in \cite{s36}
by A. and F. Shojai.

\section{SKETCH OF DEBROGLIE-BOHM-WEYL THEORY}
\renewcommand{\theequation}{2.\arabic{equation}}
\setcounter{equation}{0}

From
\cite{c7} (and references cited there) we know something about Bohmian mechanics and the quantum
potential and we go now to \cite{s33} to begin the present discussion.
In nonrelativistic deBroglie-Bohm theory the quantum potential is (cf.
\cite{c7}) ${\bf (A1)}\,\,Q=-(\hbar^2/2m)(\na^2|\Psi|/|\Psi|)$.  The
particles trajectory can be derived from Newton's law of motion in
which the quantum force $-\na Q$ is present in addition to the
classical force $-\na V$.  The enigmatic quantum behavior is attributed
here to the quantum force or quantum potential (with $\Psi$
determining a ``pilot wave" which guides the particle motion).  Setting 
$\Psi=\sqrt{\rho}exp[iS/\hbar]$ one has 
\bq\label{2.1}
\frac{\pp S}{\pp t}+\frac{|\na S|^2}{2m}+V+Q=0;\,\,\frac{\pp \rho}{\pp
t}+\na\cdot\left(\rho\frac{\na S}{m}\right)=0
\end{equation}
We follow the standard Bohmian approach here and refer to
\cite{b3,ch,c3,c7,c9,f2,f3,f7,f8} for the Bertoldi-Faraggi-Matone
development; there will be some surprising connections arising later.
The first equation in \eqref{2.1} is a Hamilton-Jacobi (HJ) equation
which is identical to Newton's law and represents an energy condition
${\bf (A2)}\,\,E=(|p|^2/2m)+V+Q$ (recall from HJ theory $-(\pp S/\pp
t)=E (=H)$ and $\na S=p$ (cf. \cite{c41}).  The second equation
represents a continuity equation for a hypothetical ensemble related to
the particle in question.  For the relativistic extension one could simply try to
generalize the relativistic energy equation ${\bf
(A3)}\,\,\eta_{\mu\nu}P^{\mu}P^{\nu}=m^2c^2$ to the form ${\bf (A4)}\,\,
\eta_{\mu\nu}P^{\mu}P^{\nu}=m^2c^2(1+{\mc Q})={\mc M}^2c^2$ where ${\bf
(A5)}\,\,{\mc Q}=
(\hbar^2/m^2c^2)(\bx |\Psi|/|\Psi|)$ and 
\bq\label{2.2}
{\mc
M}^2=m^2\left(1+\ga\frac{\bx|\Psi|}{|\Psi|}\right);\,\,\,\ga=\frac{\hbar^2}{m^2c^2}
\end{equation}
This could be derived e.g. by setting $\Psi=\sqrt{\rho}exp(iS/\hbar)$ in the
Klein-Gordon (KG) equation and separating the real and imaginary parts, leading to
the relativistic HJ equation ${\bf (A6)}\,\,\eta_{\mu\nu}\pp^{\mu}S\pp^{\nu}S={\mf
M}^2c^2$ (as in \eqref{2.1} - note $P^{\mu}=-\pp^{\mu}S$) and the continuity
equation ${\bf (A7)}\,\,\pp_{\mu}(\rho\pp^{\mu}S)=0$.  The problem of ${\mc M}^2$
not being positive definite here (i.e. tachyons) is serious however and in fact
{\bf (A4)} is not the correct equation (see e.g. \cite{ss5,s36}).  One must use the
covariant derivatives $\na_{\mu}$ in place of $\pp_{\mu}$ and for spin zero in
a curved background there results (cf. \cite{ss5,s36})
\bq\label{2.3} 
\na_{\mu}(\rho\na^{\mu}S)=0;\,\,g^{\mu\nu}\na_{\mu}S\na_{\nu}S=
{\mf M}^2c^2;
\end{equation}
To see this one must require that a correct relativistic equation of motion should
not only be Poincar\'e invariant but also it should have the correct
nonrelativistic limit.  Thus for a relativistic particle of mass ${\mf M}$
(which is a Lorentz invariant quantity) the action functional is ${\bf
(A8)}\,\,{\mf A}=\int d\gl (1/2){\mf M}(r)(dr_{\mu}/d\gl)(dr^{\nu}/d\gl)$ where
$\gl$ is any scalar parameter parametrizing the path $r_{\mu}(\gl)$ (it could e.g.
be the proper time $\tau$).  Varying the path via $r_{\mu}\to
r'_{\mu}=r_{\mu}+\gep_{\mu}$ one gets
\bq\label{2.4}
{\mf A}\to {\mf A}'={\mf A}+\gd{\mf A}={\mf A}+\int d\gl\left[\frac{dr_{\mu}}{d\gl}
\frac{d\gep^{\mu}}{d\gl}+\frac{1}{2}\frac{dr_{\mu}}{d\gl}\frac{dr^{\mu}}{d\gl}
\gep_{\nu}\pp^{\nu}{\mf M}\right]
\end{equation}
By least action the correct path satisfies ${\bf (A10)}\,\,\gd{\mf A}=0$ with fixed
boundaries so the equation of motion is ${\bf (A11)}\,\,(d/d\gl)({\mf
M}u_{\mu})=(1/2)u_{\nu}u^{\nu}\pp_{\mu}{\mf M}$ or ${\bf (A12)}\,\,{\mf
M}(du_{\mu}/d\gl)=((1/2)\eta_{\mu\nu}u_{\ga}u^{\ga}-u_{\mu}u_{\nu})\pp^{\nu}{\mf
M}$ where $u_{\mu}=dr_{\mu}/d\gl$.  Now look at the symmetries of the action
functional {\bf (A8)} via $\gl\to \gl+\gd$.  The conserved current is then
the Hamiltonian ${\mf H}=-{\mf L}+u_{\mu}(\pp{\mf L}/\pp u_{\mu})=(1/2){\mf
M}u_{\mu}u^{\mu}=E$.  This can be seen by setting $\gd{\mf A}=0$ where
\bq\label{2.5}
0=\gd{\mf A}={\mf A}'-{\mf A}=\int
d\gl\left[\frac{1}{2}u_{\mu}u^{\mu}u^{\nu}\pp_{\nu}{\mf M}+{\mf M}u_{\mu}\frac{d
u^{\mu}}{d\gl}\right]\gd
\end{equation}
which means that the integrand is zero, i.e. ${\bf (A14)}\,\,(d/d\gl)[(1/2){\mf M}
u_{\mu}u^{\mu}]=0$.  Since the proper time is defined as $c^2d\tau^2=dr_{\mu}
dr^{\mu}$ this leads to ${\bf (A15)}\,\,(d\tau/d\gl)=\sqrt{(2E/{\mf M}c^2)}$ and
the equation of motion becomes ${\bf (A16)}\,\,{\mf
M}(dv_{\mu}/d\tau)=(1/2)(c^2\eta_{\mu\nu}-v_{\mu}v_{\nu})\pp^{\nu}{\mf M}$ whre
$v_{\mu}=dr_{\mu}/d\tau$.  The nonrelativistic limit can be derived by letting the
particles velocity be ignorable with respect to light velocity.  In this limit the
proper time is identical to the time coordinate $\tau=t$ and the result is that
the $\mu=0$ component is satisfied identically via ${\bf (A17)}\,\,{\mf
M}d^2r/dt^2=-(1/2)c^2\na{\mf M}$ ($r\sim\vec{r}$).  One can write then ${\bf
(A18)}\,\,m(d^2r/dt^2)=-\na[(mc^2/2)log({\mf M}/\mu)]$ where $\mu$ is an arbitrary
mass scale.  In order to have the correct limit the term in parenthesis on the
right side should be equal to the quantum potential so ${\bf (A19)}\,\,
{\mf M}=\mu exp[-(\hbar^2/m^2c^2)(\na^2|\Psi|/|\Psi|)]$.  Thus the relativistic
quantum mass field (manifestly invariant) is ${\bf (A20)}\,\,{\mf M}=\mu
exp[(\hbar^2/2m)(\bx|\Psi|/|\Psi|)]$ and setting $\mu=m$ we get ${\bf
(A21)}\,\,{\mf M}=m exp[(\hbar^2/m^2c^2)(\bx|\Psi|/|\Psi|)]$.  If one starts with
the standard relativistic theory and goes to the nonrelativistic limit one does
not get the correct nonrelativistic equations; this is a result of 
an improper decomposition of the
wave function into its phase and norm in the KG equation (cf. also \cite{b3} for
related procedures).  One notes here also that {\bf (A21)} leads to a positive
definite mass squared.  Also from \cite{ss5} this can be extended to a many
particle version and to a curved spacetime.  In summary, for a particle in a
curved background one has (cf. \cite{s36} which we follow for the rest of this section) 
\bq\label{2.6}
\na_{\mu}(\rho\na^{\mu}S)=0;\,\,g^{\mu\nu}\na_{\mu}S\na_{\nu}S={\mf M}^2c^2;
\,\,{\mf M}^2=m^2e^{{\mf Q}};\,\,{\mf
Q}=\frac{\hbar^2}{m^2c^2}\frac{\bx_g|\Psi|}{|\Psi|}
\end{equation}
Since, following deBroglie, the quantum HJ equation in \eqref{2.6} can be written
in the form ${\bf (A22)}\,\,(m^2/{\mf M}^2)g^{\mu\nu}\na_{\mu}S\na_{\nu}S=m^2c^2$
the quantum effects are identical to a change of spacetime metric ${\bf
(A23)}\,\,g_{\mu\nu}\to\tl{g}_{\mu\nu}=({\mf M}^2/m^2)g_{\mu\nu}$ which is a
conformal transformation.  Therefore {\bf (A22)} becomes an equation ${\bf
(A24)}\,\,
\tl{g}^{\mu\nu}\tl{\na}_{\mu}S\tl{\na}_{\nu}S=m^2c^2$ where $\tl{\na}_{\mu}$
represents covariant differentiation with respect to the metric
$\tl{g}_{\mu\nu}$.  The continuity equation is then ${\bf
(A25)}\,\,\tl{g}_{\mu\nu}\tl{\na}_{\mu}(\rho\tl{\na}_{\nu}S)=0$.
The important conclusion here is that the presence of the quantum potential is
equivalent to a curved spacetime with its metric given by {\bf (A23)}.  This is a
geometrization of the quantum aspects of matter and it seems that there is a dual
aspect to the role of geometry in physics.  The spacetime geometry sometimes looks
like ``gravity" and sometimes reveals quantum behavior.  The curvature due to the
quantum potential may have a large influence on the classical contribution to the
curvature of spacetime.  The particle trajectory can now be derived from the
guidance relation via differentiation of \eqref{2.6} leading to the Newton
equations of motion
\bq\label{2.7}
{\mf M}\frac{d^2x^{\mu}}{d\tau^2}+{\mf M}\gG^{\mu}_{\nu\gk}u^{\nu}u^{\gk}=
(c^2g^{\mu\nu}-u^{\mu}u^{\nu})\na_{\nu}{\mf M}
\end{equation}
Using the conformal transformation above \eqref{2.7} reduces to the standard
geodesic equation.
\\[3mm]\indent
Now a general ``canonical" relativistic system consisting of gravity and classical matter 
(no quantum effects) is determined by the action
\bq\label{2.8}
{\mc A}=\frac{1}{2\gk}\int d^4x\sqrt{-g}{\mc R}+\int
d^4x\sqrt{-g}\frac{\hbar^2}{2m}\left(\frac{\rho}{\hbar^2}{\mc D}_{\mu}S{\mc
D}^{\mu}S-\frac{m^2}{\hbar^2}\rho\right)
\end{equation}
where $\gk=8\pi G$ and $c=1$ for convenience.  It was seen above that via
deBroglie the introduction of a quantum potential is equivalent to introducing a
conformal factor $\gO^2={\mf M}^2/m^2$ in the metric.  Hence in order to introduce
quantum effects of matter into the action \eqref{2.8} one uses this conformal
transformation to get ($1+Q\sim exp(Q)$)
\bq\label{2.9}
{\mf A}=\frac{1}{2\gk}\int d^4x\sqrt{-\bar{g}}(\bar{{\mc R}}\gO^2-6\bar{\na}_{\mu}
\gO\bar{\na}^{\mu}\gO)+
\end{equation}
$$+\int
d^4x\sqrt{-\bar{g}}\left(\frac{\rho}{m}\gO^2\bar{\na}_{\mu}S\bar{\na}^{\mu}S-
m\rho\gO^4\right)+\int
d^4x\sqrt{-\bar{g}}\gl\left[\gO^2-\left(1+\frac{\hbar^2}{m^2}
\frac{\bar{\bx}\sqrt{\rho}}{\sqrt{\rho}}\right)\right]$$
where a bar over any quantity means that it corresponds to the nonquantum regime.
Here only the first two terms of the expansion of ${\mf M}^2=m^2exp({\mf Q})$ 
in \eqref{2.6} have been used, namely ${\mf M}^2\sim m^2(1+{\mf Q})$.  No physical
change is involved in considering all the terms.  $\gl$ is a Lagrange multiplier
introduced to identify the conformal factor with its Bohmian value.  One uses
here $\bar{g}_{\mu\nu}$ to raise of lower indices and to evaluate the covariant
derivatives; the physical metric (containing the quantum effects of matter) is
$g_{\mu\nu}=\gO^2\bar{g}_{\mu\nu}$.  By variation of the action with respect to
$\bar{g}_{\mu\nu},\,\gO,\,\rho,\,S,$ and $\gl$ one arrives at the following
quantum equations of motion:
\begin{enumerate}
\item
The equation of motion for $\gO$
\bq\label{2.10}
\bar{{\mc
R}}\gO+6\bar{\bx}\gO+\frac{2\gk}{m}\rho\gO(\bar{\na}_{\mu}S\bar{\na}^{\mu}S-
2m^2\gO^2)+2\gk\gl\gO=0
\end{equation}
\item
The continuity equation for particles ${\bf
(A26)}\,\,\bar{\na}_{\mu}(\rho\gO^2\bar{\na}^{\mu}S)=0$
\item
The equations of motion for particles (here $a'\equiv\bar{a}$)
\bq\label{2.11}
(\bar{\na}_{\mu}S\bar{\na}^{\mu}S-m^2\gO^2)\gO^2\sqrt{\rho}+\frac{\hbar^2}{2m}
\left[\bx'\left(\frac{\gl}{\sqrt{\rho}}\right)-\gl\frac{\bx'\sqrt{\rho}}{\rho}\right]
=0
\end{equation}
\item
The modified Einstein equations for $\bar{g}_{\mu\nu}$
\bq\label{2.12}
\gO^2\left[\bar{{\mc R}}_{\mu\nu}-\frac{1}{2}\bar{g}_{\mu\nu}\bar{{\mc R}}\right]
-[\bar{g}_{\mu\nu}\bx'-\bar{\na}_{\mu}\bar{\na}_{\nu}]\gO^2-6\bar{\na}_{\mu}
\gO\bar{\na}_{\nu}\gO+3\bar{g}_{\mu\nu}\bar{\na}_{\ga}\gO\bar{\na}^{\ga}\gO+
\end{equation}
$$+\frac{2\gk}{m}\rho\gO^2\bar{\na}_{\mu}S\bar{\na}_{\nu}S-\frac{\gk}{m}
\rho\gO^2\bar{g}_{\mu\nu}\bar{\na}_{\ga}S\bar{\na}^{\ga}S+\gk
m\rho\gO^4\bar{g}_{\mu\nu}+$$
$$+\frac{\gk\hbar^2}{m^2}\left[\bar{\na}_{\mu}\sqrt{\rho}\bar{\na}_{\nu}\left(\frac
{\gl}{\sqrt{\rho}}\right)+\bar{\na}_{\nu}\sqrt{\rho}\bar{\na}_{\mu}\left(\frac{\gl}{\sqrt{\rho}}
\right)\right]-\frac{\gk\hbar^2}{m^2}\bar{g}_{\mu\nu}\bar{\na}_{\ga}\left[
\gl\frac{\bar{\na}^{\ga}\sqrt{\rho}}{\sqrt{\rho}}\right]=0$$
\item
The constraint equation
 ${\bf (A27)}\,\,\gO^2=1+(\hbar^2/m^2)[(\bar{\bx}\sqrt{\rho})/\sqrt{\rho}]$
\end{enumerate}
Thus the back reaction effects of the quantum factor on the background metric are
contained in these highly coupled equations.  A simpler form of \eqref{2.10} can
be obtained by taking the trace of \eqref{2.12} and using \eqref{2.10} which
produces ${\bf
(A28)}\,\,\gl=(\hbar^2/m^2)\bar{\na}_{\mu}[\gl(\bar{\na}^{\mu}\sqrt{\rho})/
\sqrt{\rho}]$.  A solution of this via perturbation methods using the small
parameter $\ga=\hbar^2/m^2$ yields the trivial solution $\gl=0$ so the above
equations reduce to
\bq\label{2.13}
\bar{\na}_{\mu}(\rho\gO^2\bar{\na}^{\mu}S)=0;\,\,\bar{\na}_{\mu}S\bar{\na}^{\mu}
S=m^2\gO^2;\,\,{\mf G}_{\mu\nu}=-\gk{\mf T}^{(m)}_{\mu\nu}-\gk{\mf
T}^{(\gO)}_{\mu\nu}
\end{equation}
where ${\mf T}_{\mu\nu}^{(m)}$ is the matter energy-momentum (EM) tensor and 
\bq\label{2.14}
\gk{\mf T}_{\mu\nu}^{(\gO)}=\frac{[g_{\mu\nu}\bx-\na_{\mu}\na_{\nu}]\gO^2}{\gO^2}
+6\frac{\na_{\mu}\gO\na_{\nu}\gO}{\go^2}-2g_{\mu\nu}\frac{\na_{\ga}\gO\na^{\ga}
\gO}{\gO^2}
\end{equation}
with ${\bf (A29)}\,\,\gO^2=1+\ga(\bar{\bx}\sqrt{\rho}/\sqrt{\rho})$.  Note that
the second relation in \eqref{2.13} is the Bohmian equation of motion and written
in terms of $g_{\mu\nu}$ it becomes $\na_{\mu}S\na^{\mu}S=m^2c^2$.
\\[3mm]\indent
In the preceeding one has tacitly assumed that there is an ensemble of quantum
particles so what about a single particle?  One translates now the quantum
potential into purely geometrical terms without reference to matter parameters 
so that the original form of the quantum potential can only be deduced after using
the field equations.  Thus the theory will work for a single particle or an
ensemble (cf. also Remark 3.3).  
Thus first ignore gravity and look at the geometrical properties of the
conformal factor given via
\bq\label{2.15}
g_{\mu\nu}=e^{4\gS}\eta_{\mu\nu};\,\,e^{4\gS}=\frac{{\mf M}^2}{m^2}=
exp\left(\ga\frac{\bx_{\eta}\sqrt{\rho}}{\sqrt{\rho}}\right)=exp\left(\ga\frac
{\bx_{\eta}\sqrt{|{\mf T}|}}{\sqrt{|{\mf T}|}}\right)
\end{equation}
where ${\mf T}$ is the trace of the EM tensor and is substituted for $\rho$ 
(true for dust).  The Einstein tensor for this metric is ${\bf (A30)}\,\,{\mf
G}_{\mu\nu}=4g_{\mu\nu}\bx_{\eta}exp(-\gS)+2exp(-2\gS)\pp_{\mu}\pp_{\nu}exp(2\gS)$.
Hence as an Ansatz one can suppose that in the presence of gravitational effects
the field equation would have a form
\bq\label{2.16}
{\mc R}_{\mu\nu}-\frac{1}{2}{\mc R}g_{\mu\nu}=\gk{\mf
T}_{\mu\nu}+4g_{\mu\nu}e^{\gS}\bx e^{-\gS}+2e^{-2\gS}\na_{\mu}\na_{\nu}e^{2\gS}
\end{equation}
This is written in a manner such that in the limit ${\mf T}_{\mu\nu}\to 0$ one
will obtain \eqref{2.15}.  Taking the trace of the last equation one gets
${\bf (A31)}\,\,-{\mc R}=\gk{\mf T}-12\bx\gS+24(\na\gS)^2$ which has the iterative
solution ${\bf (A32)}\,\,\gk{\mf T}=-{\mc R}+12\ga\bx[(\bx\sqrt{{\mc R}})/
\sqrt{{\mc R}}]$ leading to ${\bf (A33)}\,\,\gS=\ga[(\bx\sqrt{|{\mf
T}|}/\sqrt{|{\mf T}|})]\simeq \ga[(\bx\sqrt{|{\mc R}|})/\sqrt{|{\mc R}|})]$
to first order in $\ga$.
\\[3mm]\indent
One goes now to the field equations for a toy model.  First from the above one
sees that ${\mf T}$ can be replaced by ${\mc R}$ in the expression for the quantum
potential or for the conformal factor of the metric.  This is important since the
explicit reference to ensemble density is removed and the theory works for a
single particle or an ensemble.  So from \eqref{2.16} for a toy quantum gravity
theory one assumes the following field equations
\bq\label{2.17}
{\mf G}_{\mu\nu}-\gk{\mf T}_{\mu\nu}-{\mf
Z}_{\mu\nu\ga\gb}exp\left(\frac{\ga}{2}\Phi\right)\na^{\ga}\na^{\gb}exp
\left(-\frac{\ga}{2}\Phi\right)=0
\end{equation}
where ${\bf (A34)}\,\,{\mf
Z}_{\mu\nu\ga\gb}=2[g_{\mu\nu}g_{\ga\gb}-g_{\mu\ga}g_{\nu\gb}]$ and $\Phi=(\bx
\sqrt{|{\mc R}|}/\sqrt{|{\mc R}|})$.  The number 2 and the minus sign of the
second term in {\bf (A34)} are chosen so that the energy equation derived later
will be correct.  Note that the trace of \eqref{2.17} is ${\bf (A35)}\,\,
{\mc R}+\gk{\mf T}+6exp(\ga\Phi/2)\bx exp(-\ga\Phi/2)=0$ and this represents 
the connection of the Ricci scalar curvature of space time and the trace of the
matter EM tensor.  If a perturbative solution is admitted one can expand in powers
of $\ga$ to find ${\bf (A36)}\,\,{\mc R}^{(0)}=-\gk{\mf T}$ and ${\mc R}^{(1)}=
-\gk{\mf T}-6exp(\ga\Phi^0/2)\bx exp(-\ga\Phi^0/2)$ where $\Phi^{(0)}=\bx
\sqrt{|{\mf T}|}/\sqrt{|{\mf T}|}$.  The energy relation can be obtained by taking
the four divergence of the field equations and since the divergence of the
Einstein tensor is zero one obtains 
\bq\label{2.18}
\gk\na^{\nu}{\mf T}_{\mu\nu}=\ga{\mc R}_{\mu\nu}\na^{\nu}\Phi-\frac{\ga^2}{4}
\na_{\mu}(\na\Phi)^2+\frac{\ga^2}{2}\na_{\mu}\Phi\bx\Phi
\end{equation}
For a dust with ${\bf (A37)}\,\,{\mf T}_{\mu\nu}=\rho u_{\mu}u_{\nu}$ and
$u_{\mu}$ the velocity field, the conservation of mass law is ${\bf
(A38)}\,\,\na^{\nu}(\rho{\mf M} u_{\nu})=0$ so one gets to first order in $\ga$ ${\bf
(A39)}\,\,\na_{\mu}{\mf M}/{\mf M}=-(\ga/2)\na_{\mu}\Phi$ or ${\bf (A40)}\,\,{\mf
M}^2=m^2exp(-\ga\Phi)$ where $m$ is an integration constant.  This is the correct
relation of mass and quantum potential.
\\[3mm]\indent
There is then some discussion about making the conformal factor dynamical 
via a general scalar tensor action (cf. also
\cite{ss4}) and subsequently one makes both the conformal factor and the quantum
potential into dynamical fields and creates a scalar tensor theory with two scalar fields.  Thus
start with a general action
\bq\label{2.19}
{\mf A}=\int d^4x\sqrt{-g}\left[\phi{\mc R}-\go\frac{\na_{\mu}\phi\na^{\mu}\phi}{\phi}
-\frac{\na_{\mu}Q\na^{\mu} Q}{\phi}+2\gL\phi +{\mf L}_m\right]
\end{equation}
The cosmological constant generally has an interaction term with the scalar field and here
one uses an ad hoc matter Lagrangian
\bq\label{2.20}
{\mf L}_m=\frac{\rho}{m}\phi^a\na_{\mu}S\na^{\mu}S-m\rho\phi^b-\gL(1+Q)^c+\ga\rho(e^{\ell
Q}-1)
\end{equation}
(only the first two terms $1+Q$ from $exp(Q)$ are used for simplicity in the third term).
Here $a,b,c$ are constants to be fixed later and the last term is chosen 
(heuristically) in such a manner
as to have an interaction between the quantum potential field and the ensemble density (via 
the equations of motion); further the interaction is chosen so that it
vanishes in the classical limit but this is ad hoc.  Variation of the above action yields
\begin{enumerate}
\item
The scalar fields equation of motion
\bq\label{2.21}
{\mc R}+\frac{2\go}{\phi}\bx\phi-\frac{\go}{\phi^2}\na^{\mu}\phi\na_{\mu}\phi+2\gL+
\end{equation}
$$+\frac{1}{\phi^2}\na^{\mu}Q\na_{\mu}Q+\frac{a}{m}\rho\phi^{a-1}\na^{\mu}S
\na_{\mu}S-mb\rho\phi^{b-1}=0$$
\item
The quantum potential equations of motion
${\bf (A41)}\,\,(\bx Q/\phi)-(\na_{\mu}Q\na^{\mu}\phi/\phi^2)-\gL c(1+Q)^{c-1}+\ga\ell\rho
exp(\ell Q)=0$
\item
The generalized Einstein equations
\bq\label{2.22}
{\mf G}^{\mu\nu}-\gL g^{\mu\nu}=-\frac{1}{\phi}{\mf
T}^{\mu\nu}-\frac{1}{\phi}[\na^{\mu}\na^{\nu}-g^{\mu\nu}\bx]\phi+\frac{\go}{\phi^2}\na^{\mu}\phi
\na^{\nu}\phi-
\end{equation}
$$-\frac{\go}{2\phi^2}g^{\mu\nu}\na^{\ga}\phi\na_{\ga}\phi+\frac{1}{\phi^2}\na^{\mu}Q\na^{\nu}Q
-\frac{1}{2\phi^2}g^{\mu\nu}\na^{\ga}Q\na_{\ga}Q$$
\item
The continuity equation ${\bf (A42)}\,\,\na_{\mu}(\rho\phi^a\na^{\mu}S)=0$
\item
The quantum Hamilton Jacobi equation ${\bf (A43)}\,\,\na^{\mu}S\na_{\mu}S=m^2\phi^{b-a}
-\ga m\phi^{-a}(e^{\ell Q}-1)$
\end{enumerate}
In \eqref{2.21} the scalar curvature and the term $\na^{\mu}S\na_{\mu}S$ can be eliminated
using \eqref{2.22} and {\bf (A43)}; further on using the matter Lagrangian and the
definition of the EM tensor one has
\bq\label{2.23}
(2\go -3)\bx\phi =(a+1)\rho\ga(e^{\ell Q}-1)-2\gL(1+Q)^c+2\gL\phi-\frac{2}{\phi}\na_{\mu}Q
\na^{\mu}Q
\end{equation}
(where $b=a+1$).  Solving {\bf (A41)} and \eqref{2.23} with a perturbation expansion in
$\ga$ one finds
\bq\label{2.24}
Q=Q_0+\ga Q_1+\cdots;\,\,\phi=1+\ga Q_1+\cdots;\,\,\sqrt{\rho}=\sqrt{\rho_0}+
\ga\sqrt{\rho_1}+\cdots
\end{equation}
where the conformal factor is chosen to be unity at zeroth order so that as $\ga\to 0$
{\bf (A43)} goes to the classical HJ equation.  Further since by {\bf (A43)} the quantum mass
is $m^2\phi+\cdots$ the first order term in $\phi$ is chosen to be $Q_1$  (cf. \eqref{2.6}).
Also we will see that $Q_1\sim \bx\sqrt{\rho}/\sqrt{\rho}$ plus corrections which is in
accord with Q as a quantum potential field.  In any case after some computation one 
obtains ${\bf (A44)}\,\,a=2\go k,\,\,b=a+1,$ and $\ell=(1/4)(2\go k+1)=(1/4)(a+1)=b/4$ 
with
$Q_0=[1/c(2c-3)]
\{[-(2\go k+1)/2\gL]k\sqrt{\rho_0}-(2c^2-c+1)\}$ while $\rho_0$ can be
determined (cf. \cite{s36} for details).  Thus heuristically the quantum potential can be regarded as
a dynamical field and perturbatively one gets the correct dependence of quantum potential upon
density, modulo some corrective terms.
\\[3mm]\indent
One goes next to a number of examples and we only consider here the conformally flat solution
(cf. also \cite{s33}).  Thus take ${\bf (A45)}\,\,g_{\mu\nu}=exp(2\gS)\eta_{\mu\nu}$
where $\gS<< 1$.  One obtains from \eqref{2.16} ${\bf (A46)}\,\,{\mc
R}_{\mu\nu}=\eta_{\mu\nu}\bx\gS+2\pp_{\mu}
\pp_{\nu}\gS\Rightarrow {\mf G}_{\mu\nu}=2\pp_{\mu}\pp_{\nu}\gS-2\eta_{\mu\nu}\bx\gS$.
One can solve iteratively to get
\bq\label{2.25}
{\mc R}^{(0)}=-\gk{\mf T}\Rightarrow\gS^{(0)}=-\frac{\gk}{6}\bx^{-1}{\mf T};
\end{equation}
$${\mc R}^{(1)}=-\gk{\mf T}+3\ga\bx\frac{\bx\sqrt{|{\mf T}|}}{\sqrt{|{\mf T}|}}\Rightarrow
\gS^{(1)}=-\frac{\gk}{6}\bx^{-1}{\mf T}+\frac{\ga}{2}\frac{\bx\sqrt{|{\mf T}|}}
{\sqrt{|{\mf T}|}}$$
Consequently
\bq\label{2.26}
\gS=-\frac{\gk}{6}\bx^{-1}{\mf T}+\frac{\ga}{2}\frac{\bx\sqrt{|{\mf T}|}}{\sqrt{|{\mf T}|}}
+\cdots
\end{equation}
The first term is pure gravity, the second pure quantum, and the remaining terms involve
gravity-quantum interactions.  A number of impressive examples are given (cf. also
\cite{s33}). 
\\[3mm]\indent
One goes now to a generalized equivalence principle.  The gravitational effects determine
the causal structure of spacetime as long as quantum effects give its conformal structure.
This does not mean that quantum effects have nothing to do with the causal structure; they
can act on the causal structure through back reaction terms appearing in the metric field
equations.  The conformal factor of the metric is a function of the quantum potential and
the mass of a relativistic particle is a field produced by quantum corrections to the
classical mass.  One has shown that the presence of the quantum potential is equivalent to a
conformal mapping of the metric.  Thus in different conformally related frames one feels
different quantum masses and different curvatures.  In particular there are two frames with
one containing the quantum mass field and the classical metric while the other contains
the classical mass and the quantum metric.  In general frames both the spacetime metric and
the mass field have quantum properties so one can state that different conformal frames are
identical pictures of the gravitational and quantum phenomena.  We feel different quantum
forces in different conformal frames.  The question then arises of whether the
geometrization of quantum effects implies conformal invariance just as gravitational effects
imply general coordinate invariance.  One sees here that Weyl geometry provides 
additional degrees of freedom which can be identified with quantum effects and seems
to create a unified geometric framework for understanding both gravitational and quantum
forces.  Some features here are: (i) Quantum effects appear independent of any preferred
length scale.  (ii) The quantum mass of a particle is a field.  (iii) The gravitational
constant is also a field depending on the matter distribution via the quantum potential
(cf. \cite{ss4, s37}).  (iv)  A local variation of matter field distribution changes the
quantum potential acting on the geometry and alters it globally; the nonlocal character is
forced by the quantum potential (cf. \cite{s32}).

\subsection{DIRAC-WEYL ACTION}

Next (still following \cite{s36}) one goes to Weyl geometry based on the Weyl-Dirac action
\bq\label{2.27}
{\mf A}=\int d^4x\sqrt{-g}(F_{\mu\nu}F^{\mu\nu}-\gb^2\,\,{}^W{\mc
R}+(\gs+6)\gb_{;\mu}\gb^{;\mu}+ {\mf L}_{matter}
\end{equation}
Here $F_{\mu\nu}$ is the curl of the Weyl 4-vector $\phi_{\mu}$, $\gs$ is an arbitrary
constant and $\gb$ is a scalar field of weight $-1$.  The
``;" represent covariant derivatives under general coordinate and conformal transformations (Weyl
covariant derivative) defined as ${\bf (A47)}\,\,X_{;\mu}={}^W\na_{\mu}X-{\mc N}\phi_{\mu}X$
where 
${\mc N}$ is the Weyl weight of X.  The equations of motion are then
\bq\label{2.28}
{\mf G}^{\mu\nu}=-\frac{8\pi}{\gb^2}({\mf
T}^{\mu\nu}+M^{\mu\nu})+\frac{2}{\gb}(g^{\mu\nu}{}^W\na^{\ga}{}^W\na_{\ga}\gb-{}^W\na^{\mu}{}^W\na^{\nu}
\gb)+
\end{equation}
$$+\frac{1}{\gb^2}(4\na^{\mu}\gb\na^{\nu}\gb-g^{\mu\nu}\na^{\ga}\gb\na_{\ga}\gb)+\frac{\gs}
{\gb^2}(\gb^{;\mu}\gb^{;\nu}-\frac{1}{2}g^{\mu\nu}\gb^{;\ga}\gb_{;\ga});$$
$${}^W\na_{\mu}F^{\mu\nu}=\frac{1}{2}\gs(\gb^2\phi^{\mu}+\gb\na^{\mu}\gb)+4\pi J^{\mu};\,\,
{\mc
R}=-(\gs+6)\frac{{}^W\bx\gb}{\gb}+\gs\phi_{\ga}\phi^{\ga}-\gs{}^W\na^{\ga}\phi_{\ga}+\frac
{\psi}{2\gb}$$
where ${\bf
(A48)}\,\,M^{\mu\nu}=(1/4\pi)[(1/4)g^{\mu\nu}F^{\ga\gb}F_{\ga\gb}-F^{\mu}_{\ga}F^{\nu\ga}]$ 
and 
\bq\label{2.29}
8\pi{\mf T}^{\mu\nu}=\frac{1}{\sqrt{-g}}\frac{\gd\sqrt{-g}{\mf L}_{matter}}{\gd
g_{\mu\nu}};\,\,
16\pi J^{\mu}=\frac{\gd {\mf L}_{matter}}{\gd\phi_{\mu}};\,\,\psi=\frac{\gd{\mf
L}_{matter}}{\gd \gb}
\end{equation}
For the equations of motion of matter and the trace of the EM tensor one uses invariance of
the action under coordinate and gauge transformations, leading to
\bq\label{2.30}
{}^W\na_{\nu}{\mf T}^{\mu\nu}-{\mf
T}\frac{\na^{\mu}\gb}{\gb}=J_{\ga}\phi^{\ga\mu}-
\left(\phi^{\mu}+\frac{\na^{\mu}\gb}{\gb}\right){}^W\na_{\ga}J^{\ga};\,\,
16\pi{\mf T}-16\pi{}^W\na_{\mu}J^{\mu}-\gb\psi=0
\end{equation}
The first relation is a geometrical identity (Bianchi identity) and the second shows the
mutual dependence of the field equations.  Note that in the Weyl-Dirac theory the Weyl
vector does not couple to spinors so $\phi_{\mu}$ cannot be interpreted as the EM potential;
the Weyl vector is used as part of the spacetime geometry and the auxillary field (gauge field)
$\gb$ represents the quantum mass field.  The gravity fields $g_{\mu\nu}$ and $\phi_{\mu}$ and
the quantum mass field determine the spacetime geometry.   Now one constructs a Bohmian 
quantum gravity which is conformally invariant in the framework of Weyl geometry.  If the
model has mass this must be a field (since mass has non-zero Weyl weight).  The Weyl-Dirac 
action is a general Weyl invariant action as above and for simplicity now assume the matter
Lagrangian does not depend on the Weyl vector so that $J_{\mu}=0$.  The equations of motion
are then
\bq\label{2.31}
{\mf G}^{\mu\nu}=-\frac{8\pi}{\gb^2}({\mf
T}^{\mu\nu}+M^{\mu\nu})+\frac{2}{\gb}(g^{\mu\nu}{}^W\na^{\ga}{}^W\na_{\ga}\gb-{}^W\na^{\mu}{}^W\na^{\nu}
\gb)+
\end{equation}
$$+\frac{1}{\gb^2}(4\na^{\mu}\gb\na^{\nu}\gb-g^{\mu\nu}\na^{\ga}\gb\na_{\ga}\gb)+\frac{\gs}
{\gb^2}\left(\gb^{;\mu}\gb^{;\nu}-\frac{1}{2}g^{\mu\nu}\gb^{;\ga}\gb_{;\ga}\right);$$
$${}^W\na_{\nu}F^{\mu\nu}=\frac{1}{2}\gs(\gb^2\phi^{\mu}+\gb\na^{\mu}\gb);\,\,{\mc R}=
-(\gs+6)\frac{{}^W\bx\gb}{\gb}+\gs\phi_{\ga}\phi^{\ga}-\gs{}^W\na^{\ga}\phi_{\ga}
+\frac{\psi}{2\gb}$$
The symmetry conditions are ${\bf (A49)}\,\,{}^W\na_{\nu}{\mf T}^{\mu\nu}-{\mf T}
(\na^{\mu}\gb/\gb)=0$ and $16\pi{\mf T}-\gb\psi=0$ (recall ${\mf T}={\mf T}_{\mu\nu}^{\mu\nu}$).  
One notes that from \eqref{2.31}
results ${\bf (A50)}\,\,{}^W\na_{\mu}(\gb^2\phi^{\mu}+\gb\na^{\mu}\gb)=0$ 
(since $F^{\mu\nu}=0$ - cf. {\bf (B10b)}) so $\phi_{\mu}$ is
not independent of $\gb$.  To see how this is related to the Bohmian quantum theory one
introduces a quantum mass field and shows it is proportional to the Dirac field.  Thus
using \eqref{2.31} and {\bf (A49)} one has
\bq\label{2.32}
\bx\gb+\frac{1}{6}\gb{\mc R}=\frac{4\pi}{3}\frac{{\mf
T}}{\gb}+\gs\gb\phi_{\ga}\phi^{\ga}+2(\gs-6)\phi^{\gag}\na_{\gag}\gb+\frac{\gs}{\gb}
\na^{\mu}\gb\na_{\mu}\gb
\end{equation}
This can be solved iteratively via
${\bf (A51)}\,\,\gb^2=(8\pi{\mf T}/{\mc R})-\{1/[({\mc
R}/6)-\gs\phi_{\ga}\phi^{\ga}]\}\gb\bx\gb+\cdots$.  Now assuming ${\mf
T}^{\mu\nu}=\rho u^{\mu}u^{\nu}$ (dust with ${\mf T}=\rho$) we multiply {\bf (A49)} by $u_{\mu}$ and
sum to get 
${\bf (A52a)}\,\,{}^W\na_{\nu}(\rho u^{\nu})-\rho(u_{\mu}\na^{\mu}\gb/\gb)=0$.  Then put {\bf (A49)}
into {\bf (A52a)} which yields
${\bf (A52b)}\,\,
u^{\nu}{}^W\na_{\nu}u^{\mu}=(1/\gb)(g^{\mu\nu}-u^{\mu}u^{\nu})\na_{\nu}\gb$. 
To see this write (assuming $g^{\mu\nu}\na_{\nu}\gb=\na^{\mu}\gb$)
\bq\label{2.333}
{}^W\na_{\nu}(\rho
u^{\mu}u^{\nu})=u^{\mu}{}^W\na_{\nu}\rho u^{\mu}+\rho u^{\nu}{}^W\na_{\nu}u^{\mu}\Rightarrow
\end{equation}
$$\Rightarrow
u^{\mu}\left(\frac{u_{\mu}\na^{\mu}\gb}{\gb}\right)+u^{\nu}{}^W\na_{\nu}u^{\mu}-
\frac{\na^{\mu}\gb}{\gb}=0\Rightarrow u^{\nu}{}^W\na_{\nu}u^{\mu}=(1-u^{\mu}u_{\mu})\frac
{\na^{\mu}\gb}{\gb}=$$
$$(g^{\mu\nu}-u^{\mu}u_{\mu}g^{\mu\nu})\frac{\na_{\nu}\gb}{\gb}=(g^{\mu\nu}-
u^{\mu}u^{\nu})\frac{\na_{\nu}\gb}{\gb}$$
which is {\bf (A52b)}.
Then from {\bf (A51)}
\bq\label{2.322}
\gb^{2(1)}=\frac{8\pi{\mf T}}{{\mc R}};\,\,\gb^{2(2)}=\frac{8\pi{\mf T}}{{\mc R}}
\left(1-\frac{1}{({\mc R}/6)-\gs\phi_{\ga}\phi^{\ga}}\frac{\bx\sqrt{{\mf T}}}{\sqrt{{\mf
T}}}\right);\cdots
\end{equation}
Comparing with
\eqref{2.7} and \eqref{2.2} shows that we have the correct equations for the Bohmian theory
provided one identifies
\bq\label{2.33}
\gb\sim{\mf M};\,\,\frac{8\pi{\mf T}}{{\mc R}}\sim
m^2;\,\,\frac{1}{\gs\phi_{\ga}\phi^{\ga}-({\mc R}/6)}\sim \ga
\end{equation}
Thus $\gb$ is the Bohmian quantum mass field and the coupling constant $\ga$ (which depends
on $\hbar$) is also a field, related to geometrical properties of spacetime.  One notes that
the quantum effects and the length scale of the spacetime are related.  To see this suppose
one is in a gauge in which the Dirac field is constant; apply a gauge transformation to
change this to a general spacetime dependent function, i.e. ${\bf
(A53)}\,\,\gb=\gb_0\to\gb(x)=
\gb_0exp(-\Xi(x))$ via $\phi_{\mu}\to\phi_{\mu}+\pp_{\mu}\Xi$.  Thus the gauge in which the
quantum mass is constant (and the quantum force is zero) and the gauge in which the quantum
mass is spacetime dependent are related to one another via a scale change.  In particular
$\phi_{\mu}$ in the two gauges differ by $-\na_{\mu}(\gb/\gb_0)$ and since $\phi_{\mu}$ is a
part of Weyl geometry and the Dirac field represents the quantum mass one concludes that the
quantum effects are geometrized (cf. also \eqref{2.31} which shows that $\phi_{\mu}$ is not
independent of $\gb$ so the Weyl vector is determined by the quantum mass and thus the
geometrical aspect of the manifold is related to quantum effects).

\section{BACKGROUND}
\renewcommand{\theequation}{3.\arabic{equation}}
\setcounter{equation}{0}

We give now some background for the last section based on
\cite{a21,a24,a22,a23,b52,b98,b51,b53,c52,c50,c53,e2,f21,f24,f23,k50,l50,m1,m2,p10,q1,q2,r10,
s2,s4,s5,s6,s7,s31,ss3,ss4,s32,s33,s34,ss5,s35,s36,s37,s38,t2,w4,z5}.  

\subsection{WEYL GEOMETRY AND ELECTROMAGNETISM}

First we give some
background on Weyl geometry and Brans-Dicke theory following \cite{a25}; for differential geometry
we use the tensor notation of \cite{a25} and refer to e.g. \cite{bl,fe,ha,o9,or,wa,w2} for other
notation (see also \cite{w6} for an interesting variation).
One thinks of a differential manifold $M=\{U_i,\phi_i\}$ with $\phi:\,U_i\to{\bf R}^4$ and metric
$g\sim g_{ij}dx^idx^j$ satisfying $g(\pp_k,\pp_{\ell})=g_{k\ell}=<\pp_k,\pp_{\ell}>=g_{\ell k}$.
This is for the bare essentials; one can also imagine tangent vectors $X_i\sim\pp_i$ and dual
cotangent vectors $\gt^i\sim dx^i$, etc.  Given a coordinate change $\tl{x}^i=\tl{x}^i(x^j)$ a
vector $\xi^i$ transforming via ${\bf (B1)}\,\,\tl{\xi}^i=\sum \pp_i\tl{x}^j\xi^j$ is called
contravariant (e.g. $d\tl{x}^i=\sum \pp_j\tl{x}^idx^j$).
On the other hand 
$\pp\phi/\pp\tl{x}^i=\sum (\pp\phi/\pp x^j)(\pp x^j/\pp\tl{x}^i$ leads to the idea of covariant
vectors $A_j\sim\pp\phi/\pp x^j$ transforming via ${\bf (B2)}\,\,\tl{A}_i=\sum(\pp
x^j/\pp\tl{x}^i) A_j$ (i.e. $\pp/\pp \tl{x}^i\sim (\pp x^j/\pp \tl{x}^j)\pp/\pp x^j$).  Now
define  connection coefficients or Christoffel symbols via (strictly one writes
$T^{\gag}_{\;\;\ga}=g_{\ga\gb}T^{\gag\gb}$ and $T_{\ga}^{\;\;\gag}=g_{\ga\gb}T^{\gb\gag}$
which are generally different; we use that notation here but it is not used in
subsequent sections since it is unnecessary)
\bq\label{3.1}
\gG^r_{\;ki}=-\left\{\begin{array}{c}
r\\
k\,\,\,i
\end{array}\right\}=-\frac{1}{2}\sum(\pp_ig_{k\ell}+\pp_kg_{\ell i}-\pp_{\ell}g_{ik})g^{\ell r}
=\gG^r_{\;ik}
\end{equation}
(note this differs by a minus sign from some other authors).  Note also that \eqref{3.1} follows 
from equations 
\bq\label{3.2}
\pp_{\ell}g_{ik}+g_{rk}\gG^r_{\;i\ell}+g_{ir}\gG^r_{\;\ell k}=0
\end{equation}
and cyclic permutation; the basic definition of $\gG^i_{\;mj}$ is found in the transplantation law
${\bf (B3)}\,\,d\xi^i=\gG^i_{\;mj}dx^m\xi^j$.
Next
for tensors $T^{\ga}_{\gb\gag}$ define derivatives ${\bf
(B4)}\,\,T^{\ga}_{\gb\gag|k}=\pp_kT^{\ga}_{\gb\gag}$ and 
\bq\label{3.3}
T^{\ga}_{\gb\gag||\ell}=\pp_{\ell}T^{\ga}_{\gb\gag}-\gG^{\ga}_{\;\ell
s}T^s_{\gb\gag}+\gG^s_{\;\ell\gb}T^{\ga}_{s\gag}+\gG^s_{\;\ell\gag}T^{\ga}_{\gb s}
\end{equation}
In particular
covariant derivatives 
for contravariant and covariant vectors respectively are defined via
${\bf (B5)}\,\,\xi^i_{||k}=\pp_k\xi^i-\gG^i_{\;k\ell}\xi^{\ell}=\na_k\xi^i$ and $\eta_{m||\ell}
=\pp_{\ell}\eta_m+\gG^r_{\;m\ell}\eta_r=\na_{\ell}\eta_m$.
Now to describe Weyl geometry one notes first that for Riemannian geometry {\bf (B3)} holds along
with  ${\bf (B6)}\,\,\ell^2=\|\xi\|^2=g_{\ga\gb}\xi^{\ga}\xi^{\gb}$ and a scalar product formula
$\xi^{\ga}\eta_{\ga}=g_{\ga\gb}\xi^{\ga}\eta^{\gb}$.  Now however one does not demand
conservation of lengths and scalar products under affine transplantation {\bf (B3)}.  Thus assume
${\bf (B7)}\,\,d\ell=(\phi_{\gb}dx^{\gb})\ell$ where the covariant vector $\phi_{\gb}$ plays a
role analogous to $\gG^{\ga}_{\;\gb\gag}$.  Combining {\bf (B7)} with {\bf (B3)} and {\bf (B6)} one
obtains
\bq\label{3.4}
d\ell^2=2\ell^2(\phi_{\gb}dx^{\gb})=d(g_{\ga\gb}\xi^{\ga}\xi^{\gb})=
\end{equation}
$$=g_{\ga\gb|\gag}\xi^{\ga}\xi^{\gb}dx^{\gag}+g_{\ga\gb}\gG^{\ga}_{\;\rho\gag}\xi^{\rho}\xi^{\gb}
dx^{\gag}+ g_{\ga\gb}\gG^{\gb}_{\;\rho\gag}\xi^{\ga}\xi^{\rho}dx^{\gag}$$
Rearranging etc. and using {\bf (B6)} again gives ${\bf
(B8)}\,\,(g_{\ga\gb|\gag}-2g_{\ga\gb}\phi_{\gag})+g_{\gs\gb}\gG^{\gs}_{\;\ga\gag}+
g_{\gs\ga}\gG^{\gs}_{\;\gb\gag}=0$ leading to
\bq\label{3.5}
\gG^{\ga}_{\;\gb\gag}=-\left\{\begin{array}{c}
\ga\\
\gb\,\,\gag\end{array}\right\}+g^{\gs\ga}[g_{\gs\gb}\phi_{\gag}+g_{\gs\gag}\phi_{\gb}
-g_{\gb\gag}\phi_{\gs}]
\end{equation}
Thus we can prescribe the metric $g_{\ga\gb}$ and the covariant vector field $\phi_{\gag}$ and
determine by \eqref{3.5} the field of connection coefficients $\gG^{\ga}_{\;\gb\gag}$ which admits
the affine transplantation law {\bf (B3)}.  If one takes $\phi_{\gag}=0$ the Weyl geometry
reduces to Riemannian geometry.  This leads one to consider new metric tensors via ${\bf (B9)}\,\,
\hat{g}_{\ga\gb}=f(x^{\gl})g_{\ga\gb}$ and it turns out that
$(1/2)\pp log(f)/\pp x^{\gl}$ plays the role of $\phi_{\gl}$ in {\bf (B7)}.  Here {\bf (B9)} is
called a gauge transformation and the ordinary connections 
$\left\{\begin{array}{c}
\ga\\
\gb\,\,\gag\end{array}\right\}$ constructed from $g_{\ga\gb}$ are equal to the more general
connections $\hat{\gG}^{\ga}_{\;\gb\gag}$ constructed according to \eqref{3.5} from
$\hat{g}_{\ga\gb}$ and $\hat{\phi}_{\gl}=(1/2)\pp log(f)/\pp x^{\gl}$.  The generalized
differential geometry is conformal in that the ratio
\bq\label{3.6}
\frac{\xi^{\ga}\eta_{\ga}}{\|\xi\|\|\eta\|}=\frac{g_{\ga\gb}\xi^{\ga}\eta^{\gb}}{[(g_{\ga\gb}\xi^{\ga}
\xi^{\gb})(g_{\ga\gb}\eta^{\ga}\eta^{\gb})]^{1/2}}
\end{equation}
does not change under the gauge transformation {\bf (B9)}.  Again if one has a Weyl geometry
characterized by $g_{\ga\gb}$ and $\phi_{\ga}$ with connections determined by \eqref{3.5}
one may replace the geometric quantities by use of a scalar field $f$ with ${\bf (B10a)}\,\,
\hat{g}_{\ga\gb}=f(x^{\gl})g_{\ga\gb},\,\,\hat{\phi}_{\ga}=\phi_{\ga}+(1/2)(log(f)_{|\ga}$ and
$\hat{\gG}^{\ga}_{\;\gb\gag}=\gG^{\ga}_{\;\gb\gag}$ without changing the intrinsic geometric
properties of vector fields; the only change is that of local lengths of a vector via
$\hat{\ell}^2=f(x^{\gl})\ell^2$.  Note that one can reduce $\hat{\phi}_{\ga}$ to the zero vector
field if and only if $\phi_{\ga}$ is a gradient field, namely
$F_{\ga\gb}=\phi_{\ga|\gb}-\phi_{\gb|\ga}=0$ (i.e. $\phi_{\ga}=(1/2)\pp_alog(f)\equiv
\pp_{\gb}\phi_{\ga}=\pp_{\ga}\phi_{\gb}$).  In this case one has length preservation after
transplantation around an arbitrary closed curve and the vanishing of $F_{\ga\gb}$ guarantees
a choice of metric in which the Weyl geometry becomes Riemannian; thus $F_{\ga\gb}$ is an
intrinsic geometric quantity for Weyl geometry (note $F_{\ga\gb}=-F_{\gb\ga}$ and
${\bf (B10b)}\,\,\{F_{\ga\gb|\gag}\}=0$ where
$\{F_{\mu\nu|\gl}\}=F_{\mu\nu|\gl}+F_{\gl\mu|\nu}+F_{\nu\gl|\mu}$).  Similarly the concept of
covariant differentiation depends only on the idea of vector transplantation.  Indeed one can
define ${\bf (B11)}\,\,\xi^{\ga}_{||\gb}=\xi^{\ga}_{|\gb}-\gG^{\ga}_{\;\gb\gag}\xi^{\gag}$. In
Riemann geometry the curvature tensor is ${\bf
(B12)}\,\,\xi^{\ga}_{||\gb|\gag}-\xi^{\ga}_{||\gag|\gb}=R^{\ga}_{\eta\gb\gag}\xi^{\eta}$,  Hence
here we can write  ${\bf (B13)}\,\,R^{\ga}_{\;\gb\gag\gd}=-\gG^{\ga}_{\;\gb\gag|\gd}+\gG^{\ga}_
{\;\gb\gd|\gag}+\gG^{\ga}_{\;\tau\gd}\gG^{\tau}_{\;\gb\gag}-\gG^{\ga}_{\;\tau\gag}\gG^{\tau}_{\;\gb\gd}$.
Using \eqref{3.6} one then can express this in terms of $g_{\ga\gb}$ and $\phi_{\ga}$ but this is
complicated.  Equations for $R_{\gb\gd}=R^{\ga}_{\;\gb\ga\gd}$ and $R=g^{\gb\gd}R_{\gb\gd}$ are
however given in \cite{a25}.  One notes that in Weyl geometry if a vector $\xi^{\ga}$ is given,
independent of the metric,
then $\xi_{\ga}=g_{\ga\gb}\xi^{\gb}$ will depend on the metric and under a gauge transformation
one has $\hat{\xi}_{\ga}=f(x^{\gl})\xi_{\ga}$.  Hence the covariant form of a gauge invariant
contravariant vector becomes gauge dependent and one says that a tensor is of weight n if, under
a gauge transformation {\bf (B14)}, $\hat{T}^{\ga\cdots}_{\gb\cdots}=f(x^{\gl})^nT^{\ga\cdots}_
{\gb\cdots}$.  Note $\phi_{\ga}$ plays a singular role in {\bf (B10a)} and has no weight.
Similarly ${\bf (B15)}\,\,\sqrt{-\hat{g}}=f^2\sqrt{-g}$ (weight 2) and
$F^{\ga\gb}=g^{\ga\mu}g^{\gb\nu}F_{\mu\nu}$ has weight $-2$ while ${\bf (B16)}\,\,{\mf F}^{\ga\gb}
=F^{\ga\gb}\sqrt{-g}$ has weight 0 and is gauge invariant.  Similarly
$F_{\ga\gb}F^{\ga\gb}\sqrt{-g}$ is gauge invariant.
Now for Weyl's theory of electromagnetism one wants to interpret $\phi_{\ga}$ as an EM potential
and one has automatically the Maxwell equations ${\bf (B17)}\,\,\{F_{\ga\gb|\gag}\}=0$ along with
a gauge invariant complementary set ${\bf (B18)}\,\,{\mf F}^{\ga\gb}_{|\gb}={\mf s}^{\ga}$
(source equations).  These equations are gauge invariant as a natural consequence of the
geometric interpretation of the EM field.  For the interaction between the EM and
gravitational fields one sets up some field equations as indicated in \cite{a25} and the interaction
between the metric quantities and the EM fields is exhibited there. 
\\[3mm]\indent
{\bf REMARK 3.1.}
As indicated earlier in \cite{a25} $R^i_{\;jk}$ is defined with a minus sign compared with
e.g. \cite{o9,w2}.  There is also a difference in definition of the Ricci tensor which is taken
to be $G^{\gb\gd}=R^{\gb\gd}-(1/2)g^{\gb\gd}R$ in \cite{a25} with $R=R^{\gd}_{\gd}$ so that ${\bf
(B19)}\,\,G_{\mu\gag}=g_{\mu\gb}g_{\gag\gd}G^{\gb\gd}=R_{\mu\gag}-(1/2)g_{\mu\gag}R$ with
$G_{\eta}^{\gag}=R^{\eta}_{\eta}-2R\Rightarrow G_{\eta}^{\eta}=-R$ (recall $n=4$).  In \cite{o9}
the Ricci tensor is simply ${\bf (B20)}\,\,R_{\gb\mu}=R^{\ga}_{\gb\mu\ga}$ where $R^{\ga}_{\gb\mu\nu}$
is the Riemann curvature tensor and $R=R_{\eta}^{\eta}$ again.  This is similar to \cite{w2} where the
Ricci tensor is defined as $\rho_{j\ell}=R^i_{ji\ell}$.  To clarify all this we note that
${\bf (B21)}\,\,R_{\eta\gag}=R^{\ga}_{\;\eta\ga\gag}=g^{\ga\gb}R_{\gb\eta\ga\gag}=-g^{\ga\gb}
R_{\gb\eta\gag\ga}=-R^{\ga}_{\;\eta\gag\ga}$ which confirms the minus sign difference.
$\hfill\bs$

\subsection{CONFORMAL GRAVITY THEORY}

We extract here first from \cite{q2} with some embellishments (cf. also \cite{q1,s97}).
The development in \cite{q1,q2,q3,q4} is quite exhaustive and we try to capture the spirit here
(although probably providing too many details for a proper survey).  However we want to make the
treatment extensive enough to  stimulate comparison with the deBroglie-Bohm-Weyl (dBBW) theory
and to exhibit the relations between Riemannian and Weyl geometry.
Although the Jordan frame (JF) and the Einstein frame (EF) formulations of a scalar tensor
theory provide mathematically equivalent descriptions of the same physics the physical equivalence
is still under discussion (cf. \cite{f24} for a discussion of this).   There is apparently not even
agreement about which should be the physical frame and this is especially true if one allows
quantum effects to influence the metric (cf. Section 2).  The JF Lagrangian for Brans-Dicke (BD)
type theories is ${\bf (B22)}\,\,L_{BD}=(\sqrt{-g}/16\pi)(\phi R-(\go/\phi)(\na\phi)^2)$ where 
R is the Ricci scalar of the JF metric $g$, $\phi$ is the BD scalar field and $\go$ is the BD
coupling constant (a free parameter).  Under the rescaling ${\bf (B23)}\,\,\hat{g}_{ab}=\phi
g_{ab}$ and the scalar field redefinition $\hat{\phi}=log(\phi)$ the JF Lagrangian for BD
type theory is mapped into the EF Lagrangian for BD type theory, namely ${\bf (B24)}\,\,
L_E=(\sqrt{-g}/16\pi)(\hat{R}-(\go+(3/2))(\hat{\na}\hat{\phi})^2)$ where $\hat{R}$ is the
curvature scalar in terms of the EF metric $\hat{g}$.  Inserting matter involves minimal coupling
to the metric in JF theory via ${\bf (B25)}\,\,L_{JF}=(\sqrt{-g}/16\pi)(\phi R-(\go/\phi)(\na
\phi)^2)+L_{matter}$ and this is the JF formulation of BD theory.  For the EF one couples matter
minimally to the metric via ${\bf
(B26)}\,\,L_E=(\sqrt{-g}/16\pi)(\hat{R}-(\go+(3/2))(\hat{\na}\hat{\phi})^2)+L_{matter}$.  Here
the scalar field $\hat{\phi}$ is minimally coupled to curvature so the dimensional gravitational
constant G is a real constant.  Due to the minimal coupling between ordinary matter and the
spacetime metric the rest mass of any test particle m is also constant over the manifold.  This
leads to a real dimensionless gravitational coupling constant $Gm^2$ (for $\hbar=c=1$)
unlike BD theory where $Gm^2\sim\phi^{-1}$.  The equations derivable from {\bf (B26)} are
($\hat{G}_{ab}=\hat{R}_{ab}-(1/2)\hat{g}_{ab}\hat{R}$)
\bq\label{3.11}
\hat{G}_{ab}=8\pi\hat{T}_{ab}+\left(\go+\frac{3}{2}\right)(\hat{\na}_a\hat{\phi}\hat{\na}_b
\hat{\phi})-\frac{1}{2}\hat{g}_{ab}(\hat{\na}\hat{\phi})^2;\,\,\bx\hat{\phi}=0;\,\,\hat{\na}_n
\hat{T}^{na}=0
\end{equation}
Here $\hat{T}_{ab}=(2/\sqrt{-g})[\pp(\sqrt{-g}L_{matter})/\pp \hat{g}^{ab}]$.
The theory given by \eqref{3.11} is just the Einstein theory of general relativity with an
additional matter source of gravity.  For $\hat{\phi}=c$ or $\go=-3/2$ one recovers the standard
theory and this is linked with Riemannian geometry because the test particles follow the
geodesics of $\hat{g}$ via ${\bf (B27)}\,\,d^2x^a/d\hat{s}^2)=-\hat{\gG}^a_{mn}(dx^m/d\hat{s})
(dx^n/d\hat{s})$ where $\hat{\gG}^a_{bc}=(1/2)\hat{g}^{an}(\hat{g}_{bn|c}+\hat{g}_{cn|b}
-\hat{g}_{bc|n})$.  
Recall that Riemannian geometry is based on the parallel transport law
${\bf (B28)}\,\,d\xi^a=-\hat{\gag}^a_{mn}\xi^mdx^n$ with $d\hat{g}(\xi,\xi)=0$; this formulation
leads to $\hat{\gag}^a_{bc}=\hat{\gG}^a_{bc}$ (cf. Section 3.1).  Under the conformal 
transformation {\bf (B23)} the Lagrangian {\bf (B25)} is mapped into the EF Lagrangian for BD
theory, namely
\bq\label{3.12}
L_{BD}=\frac{\sqrt{-\hat{g}}}{16\pi}\left(\hat{R}-\left(\go+\frac{3}{2}\right)(\hat{\na}\hat{\phi}
)^2\right)+e^{-2\hat{\phi}}L_{matter}
\end{equation}
while {\bf (B26)} is mapped into the JF Lagrangian 
\bq\label{3.122}
L_{JF}=\frac{\sqrt{-g}}{16\pi}(\phi R+\phi^{-1}(\na\phi)^2)+\phi^2L_{matter}
\end{equation}
At the same time under {\bf (B22)} the parallel transport law {\bf (B28)} is mapped into ${\bf
(B29)}\,\,d\xi^a=-\gag^a_{mn}\xi^mdx^n$ where
$\gag^a_{bc}=\gG^a_{bc}+(1/2)\phi^{-1}(\na_b\phi\gd_c^b+\na_c\phi\gd_b^a-\na^a\phi g_{bc})$ are
the affine connections of a Weyl type manifold.  Weyl type geometry is given by the law {\bf
(B24)} along with ${\bf (B30)}\,\,dg(\xi,\xi)=\phi^{-1}dx^n\na_n\phi g(\xi,\xi)$ which is
equivalent to {\bf (B28)} with respect to the conformal transformation {\bf (B22)}.  This means
that the JF formulation of GR should be linked with a Weyl type geometry with units of measure
varying length over the manifold according to {\bf (B30)}.  In the JF GR the gravitational
constant G varies like $\phi^{-1}$ while the rest masses of material particles m vary like
$\phi^{1/2}$ (i.e. $Gm^2=c$ is preserved).  One has now two equivalent geometrical
representations of the same physical theory (according to one point of view) and we see no reason
to argue with this (see the discussion below).  The field equations of the JF theory are now
\bq\label{3.13}
G_{ab}=\frac{8\pi}{\phi}T_{ab}+\frac{\go}{\phi^2}\left(\na_a\phi\na_b\phi-\frac{1}{2}g_{ab}
g^{nm}\na_n\phi\na_m\phi\right)+\frac{1}{\phi}(\na_a\na_b\phi-g_{ab}\bx\phi)
\end{equation}
along with $\bx\phi=0$ where $T_{ab}=(2/\sqrt{-g})\pp(\sqrt{-g}\phi^2L_{matter})/\pp g^{ab}$ is
the stress energy tensor for ordinary matter in the JF.  The energy is not conserved since $\phi$
exchanges energy with the metric and matter fields and the corresponding dynamic equation is
${\bf (B31)}\,\,\na_nT^{na}=(1/2)\phi^{-1}\na^a\phi T$.  The equations of motion of an uncharged
spinless mass point acted upon by the JF metric field $g$ and by $\phi$ is
\bq\label{3.14}
\frac{d^2x^a}{ds^2}=-\gG^a_{mn}\frac{dx^m}{ds}\frac{dx^n}{ds}-\frac{1}{2}\phi^{-1}\na_n\phi
\left(\frac{dx^n}{ds}\frac{dx^a}{ds}-g^{an}\right)
\end{equation}
and this does not coincide with the geodesic equation of the JF metric.
One can also provide a new connection leading to a more canonical form of the scalar field EM
tensor in the JF (cf. \cite{s97}) and this is done in another way in \cite{q2} by
rewriting \eqref{3.13} in the form
\bq\label{3.15}
{}^{\gag}G_{ab}=\frac{8\pi}{\phi}T_{ab}+\frac{(\go+(3/2))}{\phi^2}(\na_a\phi\na_b\phi-
(1/2)g_{ab}g^{nm}\na_n\phi\na_m\phi)
\end{equation}
Again there may be energy questions but quantum input seems to render these moot (cf. Section 2).
Here one is writing \eqref{3.13} in terms of affine magnitudes in the Weyl type manifold
so that the JF Weyl manifold connections $\gag^a_{bc}$ do not coincide with the Christoffel
symbols of the JF metric $\gG^a_{bc}$; then ${}^{\gag}G_{ab}$ is given in terms of the
$\gag^a_{bc}$ instead of the $\gG^a_{bc}$.  
\\[3mm]\indent
We go next to \cite{q1,q3} and consider string connections as well (cf. also \cite{m50}).  Thus
(from \cite{q1}) first treat ${\bf (B32)}\,\,\hat{g}_{ab}=\gO^2(x)g_{ab}$ as a transformation of
units $\ul{and}$ as a conformal transformation of theory (to be further clarified via \cite{q3}).
Note e.g. that $\hat{m}=\gO^{-1}(x)m$ is not constant and consider the actions
${\bf (B33)}\,\,S_{BD}=S=\int d^4x\sqrt{-g}(\phi R-(\go/\phi)(\na \phi)^2+16\pi L_M)$
where $L_M\sim L_{matter}$ where R is the curvature scalar, $\phi$ is the BD scalar field
(the dilaton), $\go$ is the BD coupling constant, and $L_M$ is the Lagrangian of the matter
fields that are minimally coupled to the metric.  Under the change of variable $\phi\to
exp(\psi)$ BD theory can be written in the string frame ${\bf (B34)}\,\,S_S=S_1=\int d^4x\sqrt
{-g}e^{\psi}(R-\go(\na\psi)^2+16\pi e^{-\psi}L_M)$.  Dicke used the conformal transformation
{\bf (B32)} with ${\bf (B35)}\,\,\gO^2=exp(\psi)$ to rewrite the action $S_1$ (or $S_{BD}$)
in the EF, i.e. in a frame where the dilaton is minimally coupled to the curvature, namely
${\bf (B36)}\,\,S_E=S_2=\int d^4x\sqrt{-\hat{g}}(\hat{R}-(\go+(3/2))(\hat{\na}\psi)^2+16\pi
e^{-2\psi}L_M)$.  where $\hat{R}$ is the curvature scalar in terms of $\hat{g}_{ab}$ and the
matter fields are now non-minimally coupled to the dilaton $\psi$.  Another effective theory of 
gravity of BD type was proposed in \cite{m50} where there was minimal coupling of the matter
fields in the EF, namely
${\bf (B37)}\,\,S_3=\int d^4x\sqrt{-g}(R-\ga(\na\psi)^2+16\pi L_M)$ where $\ga=\go+(3/2)$.
Evidently {\bf (B37)} is just the canonical action of GR with an extra scalar (dilaton) and 
when $\ga=0$ or $\psi=c$ one recovers GR in the Einstein formulation.  Now under the conformal
transformation {\bf (B32)} - {\bf (B35)} the action {\bf (B37)} can be written in a string frame,
namely ${\bf (B38)}\,\,S_4=\int
d^4x\sqrt{-\hat{g}}e^{-\psi}(\hat{R}-(\ga-(3/2))(\hat{\na}\psi)^2+16\pi e^{-\psi}L_M)$
(different from $S_1$ in {\bf (B34)}).
The theory derivable from {\bf (B38)} is then called conformal GR or string-frame GR and
in \cite{q1} one selects from the $S_i$ those which provide a physically meaningful formulation
of the laws of gravity (i.e. a theory of gravity which is invariant under transformations of the
units of measure).  This is also elaborated below following \cite{q3}.
Now according to GR (as in {\bf (B37)} with $\ga=0$ or $\psi=c$) the structure of physical
spacetime corresponds to that of a Riemannian manifold and in general in theories with minimal
coupling of the matter to the metric are naturally linked with Riemannian manifolds.  In fact in
theories with matter of the form ${\bf (B39)}\,\,S_M=16\pi\int d^4x\sqrt{-g}L_M$ the timelike
matter particles follow free motion paths (geodesics) which are solutions of 
\bq\label{3.16}
\frac{d^2x^a}{ds^2}+\left\{\begin{array}{c}
a\\
b\,\,c\end{array}\right\}\frac{dx^m}{dx}\frac{dx^n}{ds}=0
\end{equation}
where the Christoffel symbols are $(1/2)g^{an})g_{bn|c}+g_{cn|b}-g_{bc|n}$ based on the metric
$g_{ab}$ (cf. \eqref{3.1} and \eqref{3.2}).  This situation involves the invariance of vector
lengths under parallel transport, meaning that the units of measure of the geometry are point
independent.  This situation holds then for both string frame BD theory derivable from $S_1$ and
EF GR derivable from $S_3$ where the underlying manifold is Riemannian in nature.  Under
conformal rescaling {\bf (B32)} $S_1$ and $S_3$ are mapped into their conformal versions
$S_2$ and $S_4$ respectively and at the same time manifolds of Riemannian structure are mapped
into conformally Riemannian manifolds.  Therefore in theories derivable from $S_2$ and $S_4$
we have conformally Riemannian manifolds or WIST spaces.  In particular given ${\bf (B40)}\,\,
S_M=16\pi\int d^4x\sqrt{-\hat{g}}e^{-2\psi}L_M$ the equations of motion for test particles will
be 
\bq\label{3.17}
\frac{d^2x^a}{d\hat{s}^2}+\widehat{\left\{\begin{array}{c}
a\\
m\,\,n\end{array}\right\}}\frac{dx^m}{d\hat{s}}\frac{dx^n}{d\hat{s}}-\frac{\psi_{|n}}{2}
\left(\frac{dx^n}{d\hat{s}}\frac{dx^a}{d\hat{s}}-\hat{g}^{na}\right)=0
\end{equation}
and these are conformal to \eqref{3.16}.  In this WIST geometry units of measure may change
locally.
\\[3mm]\indent
{\bf REMARK 3.2.}
This section is aimed at providing linkage points for connection to BW theory with standard
conformal gravity and trajectory equations such as \eqref{3.15}, \eqref{3.16}, and \eqref{3.17}
should be useful in this direction (see also \cite{s36}) for related examples).
$\hfill\bs$
\\[3mm]\indent
Now one looks at the group of transformations of units for length, time, and mass.  Instead of
{\bf (B32)} one considers the more general {\bf (B32)} with ${\bf
(B41)}\,\,\gO^2(x)=exp[\gs\psi(x)]$ where $\gs$ is a constant.  This can again be interpreted as a
one parameter point dependent transformation of lengths and it is interesting here to write
${\bf (B42)}\,\,\hat{\psi}=(1-\gs)\psi$ under which the basic requirements of a WIST geometry
are preserved and \eqref{3.17} is invariant under such transformations.  However this is not true
for the underlying Riemannian geometry (e.g. \eqref{3.16} is not preserved).  It can be checked
that the purely gravitational part of the actions $S_1$ and $S_4$ are invariant under {\bf (B41)},
{\bf (B42)}, and ${\bf (B43)}\,\,\hat{\ga}=[\ga/(1-\gs)^2]$ when $\gs\ne 1$.  The purely
gravitational part of the actions $S_2$ and $S_3$ is however not invariant nor is {\bf (B39)}.
On the other hand the matter action {\bf (B40)} is invariant and can be written as ${\bf
(B44)}\,\,S_M=16\pi\int d^4x\sqrt{-g}exp[2(\gs-1)\psi]$ when $\gs\ne 1$.  Therefore if one checks
the actions $S_i\,\,(i=1,\cdots,4)$ relative to invariance under {\bf (B41)}, {\bf (B42)}, and
{\bf (B43)} the only survivor is $S_4$, namely the conformal formulation of GR (or string-frame
GR).  One notes also that the set of transformations indicated is an Abelian group and
composition with parameters $\gs_1$ and $\gs_2$ involves $\gs_3=\gs_1+\gs_2-\gs_1\gs_2$ (cf.
\cite{f21}).  For reasons then spelled out more fully in \cite{q1} this group is referred to
as the group of point dependent transformations of the units of length, time, and mass;
the situation $\gs=1$ is not a member of this group; it is just a transformation allowing a jump
from one formulation of the theory to the conformal version.  The action $S_4$ is then claimed
to be the only physically meaningful formulation of the laws of gravity in this context.  This has
several implications (to be discussed below); in particular string theory may find an entry into
quantum gravity via this approach since quantum interaction will arise via the deBroglie-Bohm
quantum potential.
\\[3mm]\indent
We go next to \cite{q3} where the arguments above are developed further in an attempt to clarify
remarks in \cite{q2,q4} and establish physical equivalence among conformally related metrics.
A main point is to refute the following argument.  In canonical GR the matter couples minimally to
the metric that determines metrical relations on a Riemannian spacetime, say $\hat{g}$ (note the
switch $g\leftrightarrow \hat{g}$ here).  In this case matter particles follow the geodesics of
the metric $\hat{g}$ in Riemannian geometry and their masses are constant over the spacetime
manifold (i.e. it is the metric which matter feels so it is the physical metric).  Under the
conformal rescaling {\bf (B32)} the matter fields become non-minimally coupled to the conformal
metric $g$ and matter particles do not follow the geodesics of this last metric.  Further
it is not the metric that determines metrical relations on the manifold.  Thus although canonical
GR and its conformal image may be physically equivalent theories, nevertheless the physical metric
is that which determines metrical relations on a Riemannian spacetime and the conformal metric is
not the physical metric.  It is shown that this conclusion is wrong.  Indeed under the conformal
rescaling not only the Lagrangian is mapped into its conformal image but the spacetime geometry
itself is mapped into a conformal geometry.  In this last geometry metrical relations involve
both the conformal metric $g$ and the factor $\gO^2$ generating the transformation {\bf (B32)}. 
Hence in the conformal Lagrangian the matter fields should feel both the metric $g$ and the
scalar function $\gO$; i.e. the matter particles do not follow the geodesics of the conformal
metric alone.  The result is that under {\bf (B32)} the physical metric of the untransformed
geometry is effectively mapped into the physical metric of the conformal geometry.  Another
point involves the one parameter group of transformations of units and one shows that the only
consistent formulation of the laws of gravity (among those investigated in the paper) is the
conformal representation of general relativity.
\\[3mm]\indent
Thus, with some repetition, one looks at the effect of a conformal transformation {\bf (B32)} on
the laws of gravity and on the geometry.  The Lagrangian for canonical GR (with a scalar field)
is ${\bf
(B44)}\,\,\hat{L}_{GR}=\sqrt{-\hat{g}}(\hat{R}-\ga(\hat{\na}\hat{\phi})^2)
+16\pi\sqrt{-\hat{g}}L_M$ where $\hat{R}$ is the Ricci scalar, $(\hat{\na}\hat{\phi})^2=
\hat{g}^{mn}\hat{\phi}_{|m}\hat{\phi}_{|n}$, and $\ga\geq 0$ (again note the switch
$g\leftrightarrow \hat{g}$).  When $\phi=c$ or $\ga=0$ this is the Einstein theory.  Under
{\bf (B32)} with $\gO^2=exp(\hat{\phi})$ this becomes ${\bf (B45)}\,\,L_{GR}=\sqrt{-g}
(R-(\ga-(3/2))(\na\hat{\phi})^2)+16\pi\sqrt{-g}exp[2\hat{\phi}]L_M$.  This can be given the usual
BD form after a change of variable $\hat{\phi}\to\phi=exp(\hat{\phi})$, namely ${\bf
(B46)}\,\,L_{GR}=\sqrt{-g}(\phi R-(\ga-(3/2))[(\na\phi)^2/\phi]+16\pi\sqrt{-g}\phi^2 L_M$.
The effective gravitational constant $\hat{G}$ (set equal to 1 in {\bf (B44)}) is real and since
the matter particles follow the geodesics of $\hat{g}$ the inertial mass $\hat{m}$ is constant.
Thus the dimensionless coupling constant $\hat{G}\hat{m}^2\,\,(c=\hbar=1)$ is constant while
in conformal GR $Gm^2$ is also constant with $G\sim exp(-\hat{\phi})\sim \phi^{-1}$ and ${\bf
(B47)}\,\,m=exp[(1/2)\hat{\phi}]\hat{m}$.  On the other hand in BD theory $Gm^2\sim\phi^{-1}$.
As before one can now consider two kinds of Lagrangians for pure gravity ${\bf
(B48)}\,\,L_1=\sqrt{-g}(R-\ga(\na\phi)^2)$ and ${\bf (B49)}\,\,L_2=\sqrt{-g}(\phi
R-(\ga-(3/2))[(\na\phi)^2/\phi])$ with respect to their transformation properties under rescalings
of the units.  In particular one considers ${\bf (B50)}\,\,\tl{g}=\phi^{\gs}g_{ab}$.  Under
{\bf (B50)} $L_1$ goes to ${\bf (B51)}\,\,\tl{L}_1=\sqrt{-\tl{g}}[\phi^{\gs}\tl{R}+((3\gs-(3/2)
\gs^2)\phi^{-2-\gs}-\ga\phi^{\gs})(\tl{\na}\phi)^2]$ so the laws of gravity described by $L_1$ 
change under {\bf (B50)}.  In particular in the conformal (tilde) frame the effective
gravitational constant depends on $\phi$ due to the nonminimal coupling between the scalar field
$\phi$ and the curvature.  On the other hand $L_2$ is mapped into
\bq\label{3.19}
\tl{L}_2=\sqrt{-\tl{g}}\left[\phi^{1-\gs}\tl{R}-\frac{(\ga-(3/2)-3\gs+
(3/2)\gs^2)}{(1-\gs)^2}\phi^{\gs-1}(\tl{\na}\phi^{1-\gs})^2\right]
\end{equation}
Hence introducing a new scalar field ${\bf (B52)}\,\,\tl{\phi}=\phi^{1-\gs}$ and defining a new
parameter ${\bf (B53)}\,\,\tl{\ga}=[\ga+3\gs(\gs-2)]/(1-\gs)^2$ one can write 
\bq\label{3.20}
\tl{L}_2=\sqrt{-\tl{g}}\left[\tl{\phi}\tl{R}-\left(\ga-\frac{3}{2}\right)
\frac{(\tl{\na}\tl{\phi})^2}{\tl{\phi}}\right]
\end{equation}
Thus the Lagrangian $L_2$ is invariant in form under the conformal transformation {\bf (B50)},
the scalar field redefinition {\bf (B52)}, and the parameter transformation {\bf (B53)}.  
Such transformations are of the form indicated above with composition
$\gs_3=\gs_1+\gs_2-\gs_1\gs_2$ (cf. \cite{f21}) and the identity corresponds to $\gs=0$ with
inverse of $\gs$ being $\tl{\gs}=-\gs/(1-\gs)$ ($\gs=1$ is excluded as before - it is not a units
transformation).  Since any consistent of spacetime must be invariant under the one parameter
group of units transformation (length, time, and mass) one concludes that theories for pure
gravity described by $L_1$ are not consistent while those based on $L_2$ type Lagrangians
are consistent.  Hence e.g. canonical GR and the EF formulation of BD theory are not consistent
formulations of the laws of gravity. 
Consider now separately matter Lagrangians ${\bf (B54)}\,\,\sqrt{-g}\phi^2L_M$ and 
${\bf (B55)}\,\,\sqrt{-g}L_M$.  Here {\bf (B55)} shows minimal coupling of matter to the metric
while {\bf (B54)} has nonminimal coupling.  Under {\bf (B50)} {\bf (B54)} goes to
${\bf (B56)}\,\,\sqrt{-g}\phi^2L_M=\sqrt{-\tl{g}}\phi^{2-2\gs}L_M$ and hence considering
{\bf (B52)} one completes the demonstration that {\bf (B54)} is invariant in form under our one
parameter group of units transformations.  Unfortunately it is straightforward that {\bf (B55)}
with minimal coupling is not invariant under this group and hence BD theory (in JF formulation)
based on ${\bf (B58)}\,\,L_{BD}=L_2+16\pi\sqrt{-g}L_M$ is not yet a consistent theory of
spacetime.  The only surviving theory is the conformal GR based on {\bf (B46)}, i.e.
${\bf (B58)}\,\,L_{GR}=L_2+16\pi\phi^2L_M$ which does provide a consistent formulation of the
laws of gravity (this is BD plus nonminimal coupling).  One notes that Riemannian geometry is not
invariant under {\bf (B50)} and {\bf (B52)} so Riemannian geometry is not a consistent
formulation for the interpretation of the laws of gravity whereas Weyl geometry works.
Finally going to \cite{b98} one looks at the introduction of fields $\hat{\phi}=1+Q$ where Q is
the quantum potential in an attempt to introduce the quantum force into equations of the
form \eqref{3.17}; this is a step in the direction of consolidating BW theory with more
conventional treatments but much more is needed.
\\[3mm]\indent
{\bf REMARK 3.3}
One notes that the use of $\psi\psi^*$ automatically suggests or involves an ensemble if 
(or its square root) it is to
be interpreted as a probability density.  Thus the idea that a particle has only a probability of
being at or near x seems to mean that some paths take it there but others don't and this is
consistent with Feynman's use of path integrals for example.  This seems also to say that there
is no such thing as a particle, only a collection of versions or cloud connected to the particle
idea.  Bohmian theory on the other hand for a fixed energy gives a one parameter family of
trajectories associated to $\psi$ (see here \cite{c6} for details).  This is because the
trajectory arises from a third order differential while fixing the solution $\psi$ of the second
order stationary Schr\"odinger equation involves only two ``boundary" conditions.  As was shown in
\cite{c6} this automatically generates a Heisenberg inequality $\gD x\gD p\geq c\hbar$; i.e.
the uncertainty is built in when using the wave function $\psi$ and amazingly can be expressed by
the operator theoretical framework of quantum mechanics.  Thus a one parameter family of paths
can be associated with the use of $\psi\psi^*$ and this generates the cloud or ensemble
automatically associated with the use of $\psi$.$\hfill\bs$
\\[3mm]\indent
{\bf REMARK 3.4.}
In connection with \cite{c7} where differential calculi on fractals is mentioned it seems
promising to consider q-calculus with q related to scale, fractal dimension, and/or power laws.
This would involve a discretization but not a grid (cf. \cite{c40} for details).$\hfill\bs$

\section{CONFORMAL STRUCTURE}
\renewcommand{\theequation}{4.\arabic{equation}}
\setcounter{equation}{0}

We extract and summarize here from various sources concerning conformal geometry, QM, Bohmian
theory, etc.  First we go to \cite{au,da,sa,w3,w6} for further sketches of Weyl geometry and
will relate this to Sections 2.1 and 3.1 later (cf. also
\cite{b12,b80,b81,c51,cas,d70,ha,i2,keh,m70,m12,oli,qua,que,s6,sal,s1,s70,wa}).

\subsection{DIRAC ON WEYL GEOMETRY}

Historically of course \cite{da} takes priority and it is worthwhile to reflect on the comments
of a master craftsman.  Thus there are two papers on a new classical theory of the electron but
since this material is not essential to our needs here we omit it.  In the third paper of
\cite{da} the Dirac-Weyl action is developed (cf. also Section 2.1) and we sketch this here in
some detail.  The main point is to think of EM fields as a property of spacetime rather than
something occuring in a gravity formed spacetime.  This seems to be in the spirit of considering
a microstructure of the vacuum (or an ether) and we find it attractive.  The solution proposed by
Weyl involved a length change ${\bf (D1)}\,\,\gd\ell=\ell\gk_{\mu}\gd x^{\mu}$ under parallel
transport $x^{\mu}\to x^{\mu}+\gd x^{\mu}$.  The $\gk_{\mu}$ are field quantities occuring along
with the $g_{\mu\nu}$ in a fundamental role.  Suppose $\ell$ gets changed to $\ell'=\ell\gl(x)$
and $\ell+\gd\ell$ becomes ${\bf (D2)}\,\,\ell'+\gd\ell'=(\ell+\gd \ell)\gl(x+\gd x)=
(\ell+\gd\ell)\gl(x)+\ell\gl_{,\mu}\gd x^{\mu}$ with neglect of second order terms (here
$\gl_{,\mu}\equiv \pp\gl/\pp x^{\mu}$).  Then ${\bf (D3)}\,\,\gd\ell'=\gl\gd\ell+\ell\gl_{,\mu}
\gd x^{\mu}=\gl(\gk_{\mu}+\phi_{,\mu})\gd x^{\mu}$ where $\phi=log(\gl)$.  Hence ${\bf (D4)}\,\,
\gd\ell'=\ell'\gk'_{\mu}\gd x^{\mu}$ with $\gk'_{\mu}=\gk_{\mu}+\phi_{,\mu}.$  If the vector is
transported by parallel displacement around a small closed loop the total change in length
is ${\bf (D5)}\,\,\gd\ell=\ell F_{\mu\nu}\gd S^{\mu\nu}$ where
$F_{\mu\nu}=\gk_{\mu,\nu}-\gk_{\nu,\mu}$ and $\gd S^{\mu\nu}$ is the element of area enclosed by
the small loop.  this change is unaffected by {\bf (D4)}.  It will be seen that the field
quantities $\gk_{\mu}$ can be taken to be EM potentials, subject to the transformations {\bf
(D4)},
which correspond to no change in the geometry but a change only in the choice of artificial
standards of length.  The derived quantities $F_{\mu\nu}$ have a geometrical meaning independent
of the length standard and correspond to the EM fields.  Thus the Weyl geometry provides exactly
what is needed for describing both gravitational and EM fields in geometric terms.  There was at
first some apparent conflict with atomic standards and the theory was rejected, leaving only
the idea of gauge transformation for length standard changes.
\\[3mm]\indent
Dirac's approach however serves to help resurrect the Weyl theory; since we feel that this
theory is not perhaps sufficiently appreciated a sketch is given here (cf. however \cite{bl}).
Dirac first goes into a discussion of large numbers, e.g. $e^2/GMm$ (proton and electron masses),
$e^2/mc^2$ (age of universe), etc. and the Einsteinian theory requires that G be constant 
which seems in contradiction to $G\sim t^{-1}$ where $t$ represents the epoch time, assumed to be
increasing.  Dirac reconciles this by assuming the large numbers hypothesis (all
dimensionless large numbers are connected) and stipulating that the Einstein equations refer to
an interval $ds_E$ which is different from the interval $ds_A$ measured by atomic clocks.  Then
the objections to Weyl's theory vanish and it is assumed to refer to $ds_E$.  In this spirit then
one deals with transformations of the metric gauge under which any length such as $ds$ is
multiplied by a factor $\gl(x)$ depending on its position $x$, i.e. $ds'=\gl ds$ and a localized
quantity $Y$ may get transformed according to $Y'=\gl^nY$, in which case $Y$ is said to be of
power n and is called a co-tensor.  If $n=0$ then $Y$ is called an in-tensor and it is invariant
under gauge transformations.  The equation ${\bf (D6)}\,\,ds^2=g_{\mu\nu}dx^{\mu}dx^{\nu}$ shows
that
$g_{\mu\nu}$ is a co-tensor of power 2, since the $dx^{\mu}$ are not affected by a gauge
transformation.  Hence $g^{\mu\nu}$ is a co-tensor of power $-2$ and one writes $\sqrt{{}}$ for
$\sqrt{-g}$.  One writes $T_{:\mu}$ for the covariant derivative ($\na_{\mu}T$ would be better).
and one notes that the covariant derivative of a co-tensor is not generally a co-tensor.  However
there is a modifed covariant derivative $T_{*\mu}$ which is a co-tensor.  Consider first a scalar
$S$ of power n; then $S_{:\mu}=S_{,\mu}\equiv S_{\mu}$; under a change of gauge it transforms to
${\bf
(D7)}\,\,S'_{\mu}=(\gl^nS)_{,\mu}=\gl^nS_{\mu}+n\gl^{n-1}\gl_{\mu}S=\gl^n[S_{\mu}+
n(\gk'_{\mu}-\gk_{\mu})S]$ (via {\bf (D4)}).  Thus ${\bf
(D8)}\,\,(S_{\mu}-n\gk_{\mu}S)'=\gl^n(S_{\mu}-n\gk_{\mu}S)$ so $S_{\mu}-n\gk_{\mu}S$ is a
covector of power n and is defined to be the co-covariant derivative of S, i.e. ${\bf (D9)}\,\,
S_{*\mu}=S_{\mu}-n\gk_{\mu}S$.  To obtain the co-covariant derivative of co-vectors and co-tensors
we need a modified Christoffel symbol ${\bf
(D10)}\,\,{}^*\gG^{\ga}_{\mu\nu}=\gG^{\ga}_{\mu\nu}-g^{\ga}_{\mu}\gk_{\nu}-g^{\ga}_{\nu}\gk_{\mu}
+g_{\mu\nu}\gk^{\ga}$ (the notation $\gG^{\ga}_{\mu\nu}$ for the correct $\gG^{\ga}_{\;\mu\nu}$ is
also used in \cite{da} - cf. Section 3.1).
This is known to be invariant under gauge transformations.  Let now
$A_{\mu}$ be a co-vector of power n and form ${\bf
(D11)}\,\,A_{\mu,\nu}-{}^*\gG^{\ga}_{\mu\nu}A_{\ga}$ which is evidently a tensor since it differs
from the covariant derivative $A_{\mu:\nu}$ by a tensor and under gauge transformations one has
(cf. {\bf (D4)} where $\phi_{,\mu}=\gk'_{\mu}-\gk_{\mu}$)
\bq\label{5.1}
(A_{\mu,\nu}-{}^*\gG^{\ga}_{\mu\nu}A_{\ga})'=\gl^nA_{\mu,\nu}+n\gl^{n-1}\gl_{\nu}A_{\mu}-{}^*
\gG^{\ga}_{\mu\nu}\gl^nA_{\ga}=
\end{equation}
$$=\gl^n[A_{\mu,\nu}+n(\gk'_{\nu}-\gk_{\nu})A_{\mu}-{}^*\gG^{\ga}_{\mu\nu}A_{\ga}]$$
Thus ${\bf
(D12)}\,\,(A_{\mu,\nu}-n\gk_{\nu}A_{\mu}-{}^*\gG^{\ga}_{\mu\nu}A_{\ga})'=\gl^n[A_{\mu,\nu}
-n\gk_{\nu}A_{\mu}-{}^*\gG^{\ga}_{\mu\nu}A_{\ga}]$ so take ${\bf
(D13)}\,\,A_{\mu*\nu}=A_{\mu,\nu}-n\gk_{\nu}A_{\mu}-{}^*\gG^{\ga}_{\mu\nu}$ as the co-covariant
derivative of $A_{\ga}$.  this can be written via {\bf (D10)} as ${\bf (D14)}\,\,A_{\mu*\nu}=
A_{\mu:\nu}-(n-1)\gk_{\nu}A_{\mu}+\gk_{\mu}A_{\nu}-g_{\mu\nu}\gk^{\ga}A_{\ga}$.
Similarly for a vector $B^{\mu}$ of power n one has ${\bf (D15)}\,\,B^{\mu}_{*\nu}=B^{\mu}_{:\nu}
-(n+1)\gk_{\nu}B^{\mu}+\gk^{\mu}B_{\nu}-g^{\mu}_{\nu}\gk_{\ga}B^{\ga}$.  For a co-tensor with
various suffixes up and down one can form the co-covariant derivative via the same rules; one
notes that the co-covariant derivative always has the same power as the original.  Next observe
${\bf (D16)}\,\,(TU)_{*\gs}=T_{*\gs}U+TU_{*\gs}$ while ${\bf (D17)}\,\,g_{\mu\nu*\gs}=0$ and
$G^{\mu\nu}_{*\gs}=0$ so one can raise and lower suffixes freely in a co-tensor before carrying
out co-covariant differentiation.  Thus one can raise the $\mu$ in {\bf (D14)} giving {\bf (D15)}
with $A^{\mu}$ replacing $B^{\mu}$ and $n-2$ in place of n.  The potentials $\gk_{\mu}$ do not
form a co-vector because of the wrong transformation laws {\bf (D4)} but the $F_{\mu\nu}$
defined by {\bf (D4)} are unaffected by gauge transformations so they form an in-tensor.  One
obtains the co-covariant divergence of a co-vector $B^{\mu}$ by putting $\nu=\mu$ in {\bf (D15)}
to get ${\bf (D16)}\,\,B^{\mu}_{*\mu}=B^{\mu}_{:\mu}-(n+4)\gk_{\mu}B^{\mu}$ (for $n=-4$ this is
the ordinary covariant divergence).
\\[3mm]\indent
We list some formulas for second co-covariant derivatives now with a sketch of derivation.
Thus for a scalar of power n ${\bf (D17)}\,\,S_{*\mu*\nu}=S_{*\mu:\nu}-(n-1)\gk_{\nu}S_{*\mu}+
\gk_{\mu}S_{*\nu}-g_{\mu\nu}\gk^{\gs}S_{*\gs}$.  Putting $S_{*\mu}=S_{\mu}-n\gk_{\mu}S$ on gets
\bq\label{5.2}
S_{*\mu*\nu}=S_{\mu:\nu}-n\gk_{\mu:\nu}-n\gk_{\mu}S_{\nu}-n\gk_{\nu}(S_{\mu}-n\gk_{\mu}S)+
\gk_{\nu}S_{*\mu}+\gk_{\mu}S_{*\nu}-g_{\mu\nu}\gk^{\gs}S_{\*\gs}
\end{equation}
Now $S_{\mu:\nu}=S_{\nu:\mu}$ so ${\bf
(D18)}\,\,S_{*\mu*\nu}-S_{*\nu*\mu}=-n(\gk_{\mu:\nu}-\gk_{\nu:\mu})S=-nF_{\mu\nu}S$.
This is tedious but instructive and we continue.  Let $A_{\mu}$ be a co-vector of power n so
\bq\label{5.3}
A_{\mu*\nu*\gs}=A_{\mu*\nu:\gs}-n\gk_{\gs}A_{\mu*\nu}+(g^{\rho}_{\mu}\gk_{\gs}+g^{\rho}_{\gs}
\gk_{\mu}-g_{\mu\gs}\gk^{\rho})A_{\rho*\nu}+(g^{\rho}_{\nu}\gk_{\gs}+g^{\rho}_{\gs}\gk_{\nu}
-g_{\gs\nu}\gk^{\rho})A_{\mu*\rho}
\end{equation}
A lengthy calculation then yields ${\bf (D19)}\,\,A_{\mu*\nu*\gs}-A_{\mu*\gs*\nu}={}^*{\mf
B}_{\mu\nu\gs\rho}A^{\rho}-(n-1)F_{\nu\gs}A_{\mu}$ where
\bq\label{5.4}
{}^*{\mf
B}_{\mu\nu\gs\rho}=B_{\mu\nu\gs\rho}+g_{\rho\nu}(\gk_{\mu:\gs}+\gk_{\mu}\gk_{\gs})+g_{\mu\gs}
(k_{\rho:\nu}+\gk_{\rho}\gk_{\nu})-g_{\rho\gs}(\gk_{\mu:\nu}+\gk_{\mu}\gk_{\nu})-
\end{equation}
$$-g_{\mu\nu}(\gk_{\rho:\gs}+(\gk_{\rho}\gk_{\gs})+(g_{\rho\gs}g_{\mu\nu}-g_{\rho\nu}
g_{\mu\gs})\gk^{\ga}\gk_{\ga}$$
One can consider ${}^*{\mf B}$ as a generalized Riemann-Christoffel tensor but it does not have
the usual symmetry properties for such a tensor; however one can write ${\bf (D20)}\,\,{}^*
{\mf B}_{\mu\nu\gs\rho}={}^*B_{\mu\nu\gs\rho}+(1/2)(g_{\rho\nu}F_{\mu\gs}+g_{\mu\gs}F_{\rho\nu}
-g_{\rho\gs}F_{\mu\nu}-g_{\mu\nu}F_{\rho\gs})$ and then ${}^*B_{\mu\nu\gs\rho}$ has all the usual
symmetries, namely
\bq\label{5.5}
{}^*B_{\mu\nu\gs\rho}=-{}^*B_{\mu\gs\nu\rho}=-{}^*B_{\rho\nu\gs\mu}={}^*B_{\nu\mu\rho\gs};\,\,
{}^*B_{\mu\nu\gs\rho}+{}^*B_{\mu\gs\rho\nu}+{}^*B_{\mu\rho\nu\gs}=0
\end{equation}
Thus is is appropriate to call ${}^*B_{\mu\nu\gs\rho}$ the Riemann-Christoffel (RC) tensor for
Weyl space; it is a co-tensor of power 2.  The contracted RC tensor is ${\bf (D21)}\,\,
{}^*R_{\mu\nu}={}^*B^{\gs}_{\mu\nu\gs}=R_{\mu\nu}-\gk_{\mu:\nu}-\gk_{\nu:\mu}-g_{\mu\nu}
\gk^{\gs}_{:\gs}-2\gk_{\mu}\gk_{\nu}+2g_{\mu\nu}\gk^{\gs}\gk_{\gs}$ and is an in-tensor.  A
further contraction gives the total curvature ${\bf (D22)}\,\,{}^*R={}^*R_{\gs}^{\gs}=R
-6\gk^{\gs}_{:\gs}+6\gk^{\gs}_{\gs}$ which is a co-scalar of power -2.
\\[3mm]\indent
One gets field equations from an action principle with an in-invariant action, hence one of the
form ${\bf (D23)}\,\,I=\int\gO\sqrt{{}}d^4x$ where $\gO$ must be a co-scalar of power -4 to
compensate $\sqrt{{}}$ having power 4.  Ths usual contribution to $\gO$ from the EM field is
$(1/4)F_{\mu\nu}F^{\mu\nu}$ (of power -4 since it can be written as $F_{\mu\nu}F_{\rho\gs}
g^{\mu\rho}g^{\nu\gs}$ with F factors of power zero and $g$ factors of power -2).  One also needs
a gravitational term and the standard $-R$ could be ${}^*R$ but this has power -2 and will not
do. Weyl proposed $({}^*R)^2$ which has the correct power but seems too complicated to be
satisfactory. Here one takes ${}^*R=0$ as a constraint and puts the constraint into the Lagrangian
via $\gag {}^*R$ with $\gag$ a co-scalar field of power -2 in the form of
a Lagrange multiplier.  This leads to a scalar-tensor theory of gravitation and one can insert
other terms involving $\gag$.  For convenience one takes $\gag=-\gb^2$ with $\gb$ as the basic
field variable (co-scalar of power -1) and adds terms $k\gb^{*\gs}\gb_{*\gs}$ (co-scalar of power
-4); terms $c\gb^4$ can also be added to get ${\bf (D24)}\,\,I=\int
[(1/4)F_{\mu\nu}F^{\mu\nu}-\gb^2{}^*R+k\gb^{*\mu}\gb_{*\mu}+c\gb^4]\sqrt{{}}d^4x$ as a vacuum
action. Now $\gb^{*\mu}\gb_{*\mu}=(\gb^{\mu}+\gb\gk^{\mu})(\gb_{\mu}+\gb\gk_{\mu})$ and using
{\bf (D22)} one obtains
\bq\label{5.6}
-\gb^2{}^*R+k\gb^{*\mu}\gb_{*\mu}=-\gb^2R+k\gb^{\mu}\gb_{\mu}+(k-6)\gb^2\gk^{\mu}\gk_{\mu}
+6(\gb^2\gk^{\mu})_{:\mu}+(2k-12)\gb\gk^{\mu}\gb_{\mu}
\end{equation}
The term involving $(\gb^2\gk^{\mu})_{:\mu}$ can be discarded since its contribution to the
action density is a perfect differential, namely ${\bf (D25)}\,\,(\gb^2\gk^{\mu})_{:\mu}\sqrt{{}}
=(\gb^2\gk^{\mu}\sqrt{{}})_{,\mu}$ and for the simplest vacuum equations one chooses $k=6$ so
that {\bf (D24)} becomes ${\bf
(D26)}\,\,I=\int[(1/4)F_{\mu\nu}F^{\mu\nu}-\gb^2R+6\gb^{\mu}\gb_{\mu}+c\gb^4]\sqrt{{}} d^4x$.
Thus I no longer involves the $\gk_{\mu}$ directly but only via $F_{\mu\nu}$ and I is invariant
under transformations $\gk_{\mu}\to \gk_{\mu}+\phi_{,\mu}$ so the equations of motion that follow
from the action principle will be unaffected by such transformations (i.e. they have no physical
significance).  Now consider three kinds of transformation:
\begin{enumerate}
\item
Any transformation of coordinates.
\item
Any transformation of the metric gauge combined with the appropriate transformation of potentials
$\gk_{\mu}\to\gk_{\mu}+\phi_{,\mu}$.
\item
In the vacuum one may make a transformation of potentials as above without changing the metric
gauge or alternatively one may transform the metric gauge without changing the potentials.
This works only where there is no matter.
\end{enumerate}
For the field equations one makes small variations in all the field quantities $g_{\mu\nu},\,\,
\gk_{\mu},$ and $\gb$, calculates the change in I and sets it equal to zero.  Thus write
${\bf (D27)}\,\,\gd I=\int [(1/2)P^{\mu\nu}\gd g_{\mu\nu}+Q^{\mu}\gd\gk_{\mu}+S\gd\gb)\sqrt{{}}
d^4x$ and drop the $c\gb^4\sqrt{{}}$ term since it is probably only of interest for cosmological
purposes.  One has ${\bf (D28)}\,\,\gd[(1/4)F_{\mu\nu}F^{\mu\nu}\sqrt{{}}]=(1/2)E^{\mu\nu}
\sqrt{{}}\gd g_{\mu\nu}-J^{\mu}\sqrt{{}}\gd\gk_{\mu}$ with neglect of a perfect differential.
Here $E^{\mu\nu}$ is the EM stress tensor ${\bf (D29)}\,\,E^{\mu\nu}=(1/4)g^{\mu\nu}F^{\ga\gb}
F_{\ga\gb}-F^{\mu\ga}F^{\nu}_{\ga}$ and $J^{\mu}$ is the charge current vector ${\bf (D30)}\,\,
F^{\mu}=F^{\mu\nu}_{:\nu}=\sqrt{{}}^{-1}(F^{\mu\nu}\sqrt{{}})_{,\nu}$.
Considerable calculation and neglect of perfect differentials leads finally to
\bq\label{5.7}
P^{\mu\nu}=E^{\mu\nu}+\gb^2[2R^{\mu\nu}-g^{\mu\nu}R]-4g^{\mu\nu}\gb\gb^{\rho}_{:\rho}+
4\gb\gb^{\mu:\nu}+2g^{\mu\nu}\gb^{\gs}\gb_{\gs}-8\gb^{\mu}\gb^{\nu};
\end{equation}
$$Q^{\mu}=-J^{\mu};\,\,S=-2\gb R-12\gb^{\mu}_{:\mu}$$
and the field equations for the vacuum are ${\bf (D31)}\,\,P^{\mu\nu}=0,\,\,Q^{\mu}=0,$ and
$S=0$.  These are not all independent since ${\bf
(D32)}\,\,P^{\gs}_{\gs}=-2\gb^2R-12\gb\gb^{\gs}_{:\gs}=\gb S$ so the S equation is a consequence
of the P equations.  If one omits the EM term from the action it becomes the same as the
Brans-Dicke action except that the latter allows an arbitrary value for k; with $k\ne 6$ the
vacuum equations are independent so the BD theory has one more vacuum field equation, namely
$\bx(\gb^2)=0$.
\\[3mm]\indent
Now the action integral is invariant under transformations of the coordinate sysem and
transformations of gauge; each of these leads to a conservation law connecting the quantities
$P^{\mu\nu},\,Q^{\mu},S$ defined via {\bf (D27)}.  For coordinate transformations $x^{\mu}\to
x^{\mu}+b^{\mu}$ one gets
\bq\label{5.8}
-\gd g_{\mu\nu}=g_{\mu\gs}b^{\gs}_{,\nu}+g_{\nu\gs}b^{\gs}_{,\mu}+g_{\mu\nu,\gs}b^{\gs};\,\,
-\gd\gb=\gb_{\gs}b^{\gs};\,\,-\gd\gk_{\mu}=\gk_{\gs}b^{\gs}_{,\mu}+\gk_{\mu,\gs}b^{\gs}
\end{equation}
Putting these variations in {\bf (D27)} yields
\bq\label{5.9}
\gd
I=-\int[(1/2)P^{\mu\nu}(g_{\mu\gs}b^{\gs}_{,\nu}+g_{\nu\gs}b^{\gs}_{,\mu}+g_{\mu\nu,\gs}b^{\gs})+Q^{\mu}
(\gk_{\gs}b^{\gs}_{,\mu}+\gk_{\mu,\gs}b^{\gs})+S\gb_{\gs}b^{\gs}]\sqrt{{}}d^4x=
\end{equation}
$$=\int[(P^{\mu}_{\gs}\sqrt{{}})_{,\mu}-(1/2)P^{\mu\nu}g_{\mu\nu,\gs}\sqrt{{}}+
(Q^{\mu}\gk_{\gs}\sqrt{{}})_{,\mu}-Q^{\mu}\gk_{\mu,\gs}\sqrt{{}}-
S\gb_{\gs}\sqrt{{}}]b^{\gs}d^4x$$
This $\gd I$ vanishes for arbitrary $b^{\gs}$ so one puts the coefficient of $b^{\gs}$ equal to 
zero; using ${\bf (D33)}\,\,(P^{\mu}_{\gs}\sqrt{{}})_{,\mu}-(1/2)P^{\mu\nu}g_{\mu\nu,\gs}\sqrt{{}}
=P^{\mu}_{\gs:\mu}\sqrt{{}}$ and ${\bf (D34)}\,\,(Q^{\mu}\gk_{\gs}\sqrt{{}})_{,\mu}=\gk_{\gs}
Q^{\mu}_{:\mu}\sqrt{{}}+\gk_{\gs,\mu}Q^{\mu}\sqrt{{}}$ this reduces to ${\bf (D35)}\,\,
P^{\mu}_{\gs:\mu}+\gk_{\gs}Q^{\mu}_{:\mu}+F_{\gs\mu}Q^{\mu}-S\gb_{\gs}=0$.  Next consider a small
transformation in gauge ${\bf (D36)}\,\,\gd g_{\mu\nu}=2\gl g_{\mu\nu},\,\,\gd\gb=-\gl\gb,$ and
$\gd\gk_{\mu}=[log(1+\gl)]_{,\mu}=\gl_{\mu}$.  Putting this in {\bf (D27)} yields
\bq\label{5.10}
\gd I=\int[P^{\mu\nu}\gl g_{\mu\nu}+Q^{\mu}\gl_{\mu}-S\gl\gb]\sqrt{{}}d^4x=\int
[P^{\mu}_{\mu}\sqrt{{}}-(Q^{\mu}\sqrt{{}})_{,\mu}-S\gb\sqrt{{}}]\gl d^4x
\end{equation}
Putting the coefficient of $\gl$ equal to zero gives ${\bf
(D27)}\,\,P^{\mu}_{\mu}-Q^{\mu}_{:\mu}-S\gb=0$; here {\bf (D35)} and {\bf (D37)} are the
conservation laws.  For the vacuum one sees that {\bf (D37)} is the same as {\bf (D32)} since
$Q^{\mu}_{:\mu}=0$ from \eqref{5.7}; also {\bf (D35)} reduces to ${\bf (D38)}\,\,
P^{\mu}_{\gs:\mu}+F_{\gs\mu}Q^{\mu}-\gb^{-1}\gb_{\gs}P^{\mu}_{\mu}=0$ which may be considered as
a generalization of the Bianchi identities.  The conservation laws {\bf (D35)} and {\bf (D37)}
hold more generally than for the vacuum, namely whenever the action integral can be constructed
from the field variables $g_{\mu\nu},\,\gk_{\mu},\,\gb$ alone.
\\[3mm]\indent
Now let the coordinates of a particle be $z^{\mu}$, functions of the proper time s measured along
its world line.  Put $dz^{\mu}/ds=v^{\mu}$ for velocity so $v_{\mu}v^{\mu}=1$ and $v^{\mu}$ is a
co-vector of power -1.  One adds to the action the further terms ${\bf (D39))}\,\,I_1=-m\int \gb
ds$ and $I_2=e\int \gb^{-1}\gb_{*\mu}v^{\mu}ds$ (m and e being constants).  Then these terms are
in-invariants with ${\bf (D40)}\,\,I_2=e\int (\gb^{-1}\gb_{\mu}+\gk_{\mu})v^{\mu}ds=e\int
[(d/ds)(log(\gb))+\gk_{\mu}v^{\mu}]ds$ and the first term contributes nothing to the action
principle.  Thus $I_2=e\int \gk_{\mu}v^{\mu}ds$ which is unchanged when $\gk_{\mu}\to
\gk_{\mu}+\phi_{,\mu}$ since the extra term is $e\int (d\phi/ds)ds$.  Thus for a particle with
action $I_1+I_2$ the transformations (3) above are still possible.  Now some calculation yields
\bq\label{5.11}
m[g_{\mu\gs}d(\gb v^{\mu})/ds+\gb\gG_{\gs\mu\nu}v^{\mu}v^{\nu}-\gb_{\gs}]=-ev^{\mu}F_{\mu\gs}
\equiv m[d(\gb v^{\mu})/ds+\gG^{\mu}_{\rho\gs}v^{\rho}v^{\gs}-\gb^{\mu}]=eF^{\mu\nu}v_{\nu}
\end{equation}
This is the equation of motion for a particle of mass m and charge e; if $e=0$ it could be called
an in-geodesic.  If one works with the Einstein gauge then the case $e=0$ gives the usual
geodesic equation.  Next one considers the influence the of particle on the field and this is done
by generating a dust of particles and a continuous fluid leading to an equation
\bq\label{star}
\rho[(\gb v^{\mu})_{,\nu}+\gG^{\mu}_{\ga\gs}v^{\ga}v^{\gs}-\gb^{\mu}]=\gs ^{\mu\nu}
\end{equation} 
where $\rho$ and $\gs$ refer to mass and charge density respectively.

\subsection{THE SCHR\"ODINGER EQUATION IN WEYL SPACE}

We go now to Santamato \cite{sa} and derive the SE from classical mechanics in Weyl space
(cf. also \cite{au,cas,sb}).
The idea is to relate the quantum force (arising from the quantum potential) to geometrical
properties of spacetime; the Klein-Gordon (KG) equation is also treated in this spirit.
One wants to show how geometry acts as a guidance field for matter (as in general relativity).
Initial positions are assumed random (as in the Madelung approach) and thus the theory is really
describing the motion of an ensemble.  Thus assume that the particle motion is given by some
random process $q^i(t,\go)$ in a manifold M (where $\go$ is the sample space tag) whose
probability density $\rho(q,t)$ exists and is properly normalizable.  Assume that the process
$q^i(t,\go)$ is the solution of differential equations ${\bf (D41)}\,\,\dot{q}^i(t,\go)=
(dq^i/dt)(t,\go)=v^i(q(t,\go),t)$ with random initial conditions $q^i(t_0,\go)=q_0^i(\go)$.
Once the joint distribution of the random variables $q_0^i(\go)$ is given the process
$q^i(t,\go)$ is uniquely determined by {\bf (D41)}.  One knows that in this situation 
${\bf (D42)}\,\,\pp_t\rho+\pp_i(\rho v^i)=0$ with initial Cauchy data $\rho(q,t)=\rho_0(q)$.
The natural origin of $v^i$ arises via a least action principle based on a Lagrangian
$L(q,\dot{q},t)$ with
\bq\label{5.12}
L^*(q,\dot{q},t)=L(q,\dot{q},t)-\Phi(q,\dot{q},t);\,\,\Phi=\frac{dS}{dt}=\pp_tS+\dot{q}^i\pp_iS
\end{equation}
Then $v^i(q,t)$ arises by minimizing ${\bf
(D43)}\,\,I(t_0,t_1)=E[\int_{t_0}^{t_1}L^*(q(t,\go),\dot{q}(t,\go),t)dt]$ where $t_0,\,t_1$ are
arbitrary and E denotes the expectation (cf. \cite{c7,n9,n15,n6} for stochastic ideas). 
The minimum is to be achieved over the class of all
random motions $q^i(t,\go)$ obeying {\bf (D41)} with arbitrarily varied velocity field $v^i(q,t)$
but having common initial values.  One proves first
\bq\label{5.13}
\pp_tS+H(q,\na S,t)=0;\,\,v^i(q,t)=\frac{\pp H}{\pp p_i}(q,\na S(q,t),t)
\end{equation}
Thus the value of I in {\bf (D43)} along the random curve $q^i(t,q_0(\go))$ is ${\bf (D44)}\,\,
I(t_1,t_0,\go)=\int_{t_0}^{t_1}L^*(q(,q_0(\go)),\dot{q}(t,q_0(\go)),t)dt$.  Let
$\mu(q_0)$ denote the joint probability density of the random variables $q^i_0(\go)$ and then the
expectation value of the random integral is
\bq\label{5.14}
I(t_1,t_0)=E[I(t_1,t_0,\go)]=\int_{{\bf
R}^n}\int_{t_0}^{t_1}\mu(q_0)L^*(q(t,q_0),\dot{q}(t,q_0),t)d^nq_0dt
\end{equation}
Standard variational methods give then
\bq\label{5.15}
\gd I=\int_{{\bf R}^n}d^nq_0\mu(_0)\left[\frac{\pp L^*}{\pp\dot{q}^i}(q(t_1,q_0),\pp_tq(t_1,q_0),
t)\gd q^i(t_1,q_0)-\right.
\end{equation}
$$-\left.\int_{t_0}^{t_1}dt\left(\frac{\pp}{\pp t}\frac{\pp
L^*}{\pp\dot{q}^i}(q(t,q_0),\pp_tq)t,q_0),t)-\frac{\pp L^*}{\pp
q^i}(q(t,q_0),\pp_tq(t,q_0),t)\right)\gd q^i(t,q_0)\right]$$
where one uses the fact that $\mu(q_0)$ is independent of time and $\gd q^i(t_0,q_0)=0$ (recall
common initial data is assumed).  Therefore ${\bf (D45)}\,\,(\pp L^*/\pp
\dot{q}^i)(q(t,q_0),\pp_tq(t,q_0),t)=0$ and 
\bq\label{5.16}
\frac{\pp}{\pp t}\frac{\pp L^*}{\pp\dot{q}^i}(q(t,q_0),\pp_tq(t,q_0,t)-\frac{\pp L^*}{\pp q^i}
(q(t,q_0),\pp_tq(t,q_0),t)=0
\end{equation}
are the necessary conditions for obtaining a minimum of I.  Conditions \eqref{5.16} are the usual
Euler-Lagrange equations whereas {\bf (D45)} is a consequence of the fact that in the most
general case one must retain varied motions with $\gd q^i(t_1,q_0)$ different from zero at the
final time $t_1$.  Note that since $L^*$ differs from L by a total time derivative one can safely
replace $L^*$ by L in \eqref{5.16} and putting\eqref{5.12} into {\bf (D45)} one obtains the
classical equations ${\bf (D46)}\,\,p_i=(\pp L/\pp\dot{q}^i)(q(t,q_0),\dot{q}(t,q_0),t)=\pp_i S
(q(t,q_0),t)$.  It is known now that if ${\bf
(D47)}\,\,det[(\pp^2L/\pp\dot{q}^i\pp\dot{q}^j]\ne 0$ then the second equation in \eqref{5.13}
is a consequence of the gradient condition {\bf (D46)} and of the definition of the Hamiltonian
function $H(q,p,t)=p_i\dot{q}^i-L$.  Moreover \eqref{5.16} and {\bf (D46)} entrain the HJ
equation in \eqref{5.13}.  In order to show that the average action integral \eqref{5.14}
actually gives a minimum one needs $\gd^2I>0$ but this is not necessary for Lagrangians whose
Hamiltonian H has the form
\bq\label{5.17}
H_C(q,p,t)=\frac{1}{2m}g^{ik}(p_i-A_i)(p_k-A_k)+V
\end{equation}
with arbitrary fields $A_i$ and V (particle of mass m in an EM field A) which is the form for
nonrelativistic applications; given positive definite $g_{ik}$ such Hamiltonians involve
sufficiency conditions ${\bf (D48)}\,\,det[\pp^2L/\pp\dot{q}^i\pp\dot{q}^k]=mg>0$.  Finally
\eqref{5.16} with $L^*$ replaced by L) shows that along particle trajectories the EL equations
are satisfied, i.e. the particle undergoes a classical motion with probability one.  Notice here
that in \eqref{5.13} no explicit mention of generalized momenta is made; one is dealing with a
random motion entirely based on position.  Moreover the minimum principle {\bf (D43)}
defines a 1-1 correspondence between solutions $S(q,t)$ in \eqref{5.13} and minimizing random
motions $q^i(t,\go)$.  Provided $v^i$ is given via \eqref{5.13} the particle undergoes a
classical motion with probability one.  Thus once the Lagrangian L or equivalently the
Hamiltonian H is given {\bf (D42)} and \eqref{5.13} uniquely determine the stochastic process
$q^i(t,\go)$.  Now suppose that some geometric structure is given on M so that the notion
of scalar curvature $R(q,t)$ of M is meaningful.  Then we assume (ad hoc) that the actual
Lagrangian is ${\bf (D49)}\,\,L(q,\dot{q},t)=L_C(q,\dot{q},t)+\gag(\hbar^2/m)R(q,t)$ where
${\bf (D50)}\,\,\gag=(1/6)(n-2)/ (n-1)$ with $n=dim(M)$.  Since both $L_C$ and R are independent
of $\hbar$ we have $L\to L_C$ as $\hbar\to 0$.
\\[3mm]\indent
Now for a Riemannian geometry ${\bf (D51)}\,\,ds^2=g_{ik}(q)dq^idq^k$ it is standard that in a
transplantation
$q^i\to q^i+\gd q^i$ one has ${\bf (D52)}\,\,\gd A^i=\gG^i_{k\ell}A^{\ell}dq^k$.
Here however it is assumed that for $\ell=(g_{ik}A^iA^k)^{1/2}$ one has ${\bf
(D53)}\,\,\gd\ell=\ell\phi_kdq^k$ where the $\phi_k$ are covariant components of an arbitrary
vector of M (Weyl geometry).  For a different perspective we review the material on Weyl
geometry in \cite{sa}.  Thus the actual affine connections $\gG^i_{k\ell}$ can be found by
comparing {\bf (D53)} with $\gd\ell^2=\gd(g_{ik}A^iA^k)$ and using {\bf (D52)}.  A little linear
algebra gives then 
\bq\label{5.18}
\gG^i_{k\ell}=-\left\{\begin{array}{c}
i\\
k\,\,\ell\end{array}\right\}+g^{im}(g_{mk}\phi_{\ell}+g_{m\ell}\phi_k-g_{k\ell}\phi_m)
\end{equation}
(again in \cite{sa} the notation $\gG^i_{k\ell}$ is used in place of $\gG^i_{\;k\ell}$ - cf.
Section 3.1).
Thus we may prescribe the metric tensor $g_{ik}$ and $\phi_i$ and determine via \eqref{5.18}
the connection coefficients.  Note that $\gG^i_{k\ell}=\gG^i_{\ell k}$ and for $\phi_i=0$ one
has Riemannian geometry.  Covariant derivatives are defined for contravariant $A^k$ via
${\bf (D54)}\,\,A^k_{,\i}=\pp_iA^k-\gG^{k\ell}A^{\ell}$ and for covariant $A_k$ via
${\bf (D55)}\,\,A_{k,i}=\pp_iA_k+\gG^{\ell}_{ki}A_{\ell}$ (where $S_{,i}=\pp_iS$).  Note Ricci's
lemma  no longer holds (i.e. $g_{ik,\ell}\ne 0$) so covariant differentiation and operations of
raising or lowering indices do not commute.  The curvature tensor $R^i_{k\ell m}$ in Weyl
geometry is introduced via ${\bf (D56)}\,\,A^i_{,k,\ell}-A^i_{,\ell,k}=F^i_{mk\ell}A^m$
from which arises the standard formula of Riemannian geometry ${\bf (D57)}\,\,R^i_{mk\ell}=
-\pp_{\ell}\gG^i_{mk}+\pp_k\gG^i_{m\ell}+\gG^i_{n\ell}\gG^n_{mk}-\gG^i_{nk}\gG^n_{m\ell}$
where\eqref{5.18} must be used in place of the Christoffel symbols.  The tensor $R^i_{mk\ell}$
obeys the same symmetry relations as the curvature tensor of Riemann geometry as well as the 
Bianchi identity.  The Ricci symmetric tensor $R_{ik}$ and the scalar curvature R are defined by
the same formulas also, viz. $R_{ik}=R^{\ell}_{i\ell k}$ and $R=g^{ik}R_{ik}$.  For completeness
one derives here ${\bf
(D58)}\,\,R=\dot{R}+(n-1)[(n-2)\phi_i\phi^i-2(1/\sqrt{g})\pp_i(\sqrt{g}\phi^i)]$ where 
$\dot{R}$ is the Riemannian curvature built by the Christoffel symbols.  Thus from \eqref{5.18}
one obtains
\bq\label{5.19}
g^{k\ell}\gG^i_{k\ell}=-g^{k\ell}\left\{\begin{array}{c}
i\\
k\,\,\ell\end{array}\right\}-(n-2)\phi^i;\,\,\gG^i_{k\ell}=-\left\{\begin{array}{c}
i\\
k\,\,\ell\end{array}\right\}+n\phi_k
\end{equation}
Since the form of a scalar is independent of the coordinate system used one may compute R in a
geodesic system where the Christoffel symbols and all $\pp_{\ell}g_{ik}$ vanish; then \eqref{5.18}
reduces to ${\bf (D59)}\,\,\gG^i_{k\ell}=\phi_k\gk^i_{\ell}+\phi_{\ell}\gd^i_k-g_{k\ell}\phi^i$.
Hence ${\bf (D60)}\,\,R=-g^{km}\pp_m\gG^i_{k\ell}+\pp_i(g^{k\ell}\gG^i_{k\ell})+g^{\ell
m}\gG^i_{n\ell}\gG^n_{mi}-g^{m\ell}\gG^i_{n\ell}\gG^n_{m\ell}$.  Further from {\bf (D59)}
one has ${\bf (D61)}\,\,g^{\ell m}\gG^i_{n\ell}\gG^n_{mi}=-(n-2)(\phi_k\phi^k)$ at the point in
consideration.  Putting all this in {\bf (D60)} one arrives at ${\bf (D62)}\,\,
R=\dot{R}+(n-1)(n-2)(\phi_k\phi^k)-2(n-1)\pp_k\phi^k$ which becomes {\bf (D58)} in covariant form.
Now the geometry is to be derived from physical principles so the $\phi_i$ cannot be arbitrary
but must be obtained by the same averaged least action principle {\bf (D43)} giving the motion of
the particle.  The minimum in {\bf (D43)} is to be evaluated now with respect to the class of all
Weyl geometries having arbitrarily varied gauge vectors but fixed metric tensor.  Note that once 
{\bf (D49)} is inserted in \eqref{5.12} the only term in {\bf (D43)} containing the gauge vector
is the curvature term.  Then observing that $\gag>0$ when $n\geq 3$ the minimum principle {\bf
(D43)} may be reduced to the simpler form ${\bf (D63)}\,\,E[R(q(t,\go),t)]=min$ where only the
gauge vectors $\phi_i$ are varied.  Using {\bf (D58)} this is easily done.  First a little
argument shows that $\hat{\rho}(q,t)=\rho(q,t)/\sqrt{g}$ transforms as a scalar in a coordinate
change and this will be called the scalar probability density of the random motion of the
particle.  Starting from {\bf (D42)} a manifestly covariant equation for $\hat{\rho}$ is found to
be ${\bf (D65)}\,\,\pp_t\hat{\rho}+(1/\sqrt{g})\pp_i(\sqrt{g}v^i\hat{\rho})=0$.  Now return to
the minimum problem {\bf (D63)}; from {\bf (D58)} and {\bf (D64)} one obtains
\bq\label{5.20}
E[R(q(t,\go),t)]=E[\dot{R}(q(t,\go),t)]+(n-1)\int_M[(n-2)\phi_i\phi^i-2(1/\sqrt{g})
\pp_i(\sqrt{g}\phi^i)]\hat{\rho}(q,t)\sqrt{g}d^nq
\end{equation}
Assuming fields go to 0 rapidly enough on $\pp M$ and integrating by parts one gets then
\bq\label{5.21}
E[R]=E[\dot{R}]-\frac{n-1}{n-2}E[g^{ik}\pp_i(log(\hat{\rho})\pp_k(log(\hat{\rho})]+
\end{equation}
$$+\frac{n-1}{n-2}E\{g^{ik}[(n-2)\phi_i+\pp_i(log(\hat{\rho})][(n-2)\phi_k+\pp_k(log(\hat{\rho})]\}
$$
Since the first two terms on the right are independent of the gauge vector and $g^{ik}$ is
positive definite $E[R]$ will be a minimum when ${\bf
(D66)}\,\,\phi_i(q,t)=-[1/(n-2)]\pp_i[log(\hat{\rho})(q,t)]$.  This shows that the geometric
properties of space are indeed affected by the presence of the particle and in turn the alteration
of geometry acts on the particle through the quantum force $f_i=\gag(\hbar^2/m)\pp_iR$ which
according to {\bf (D58)} depends on the gauge vector and its derivatives.  It is this peculiar
feedback between the geometry of space and the motion of the particle which produces quantum
effects.
\\[3mm]\indent
In this spirit one goes next to a geometrical derivation of the SE.  Thus inserting {\bf (D66)}
into {\bf (D58)} one gets ${\bf
(D67)}\,\,R=\dot{R}+(1/2\gag\sqrt{\hat{\rho}})[1/\sqrt{g})\pp_i(\sqrt{g}g^{ik}\pp_k\sqrt{\rho})]$
where the value {\bf (D50)} for $\gag$ has been used.  On the other hand the HJ equation 
\eqref{5.12} can be written as ${\bf (D68)}\,\,\pp_tS+H_C(q,\na S,t)-\gag(\hbar^2/m)R=0$
where {\bf (D49)} has been used.  When {\bf (D67)} is introduced into {\bf (D68)} the HJ equation
{\bf (D76)} and the continuity equation {\bf (D65)}, with velocity field biven by \eqref{5.13},
form a set of two nonlinear PDE which are coupled by the curvature of space.  Therefore self
consistent random motions of the particle (i.e. random motions compatible with {\bf (D60)})
are obtained by solving {\bf (D65)} and {\bf (D68)} simultaneously.  For every pair of solutions
$S(q,t,\hat{\rho}(q,t))$ one gets a possible random motion for the particle whose invariant
probability density is $\hat{\rho}$.  The present approach is so different from traditional QM
that a proof of equivalence is needed and this is only done for Hamiltonians of the form 
\eqref{5.17} (which is not very restrictive).  The HJ equation {\bf (D69)} corresponding to
\eqref{5.17} is
\bq\label{5.22}
\pp_tS+\frac{1}{2m}g^{ik}(\pp_iS-A_i)(\pp_kS-A_k)+V-\gag\frac{\hbar^2}{m}R=0
\end{equation}
with R given by {\bf (D67)}.  Moreover using \eqref{5.13} as well as \eqref{5.17} the continuity
equation {\bf (D65)} becomes ${\bf
(D69)}\,\,\pp_t\hat{\rho}+(1/m\sqrt{g})\pp_i[\hat{\rho}\sqrt{g}g^{ik}(\pp_kS-A_k)]=0$.
Owing to {\bf (D67)} \eqref{5.22} and {\bf (D69)} form a set of two nonlinear PDE which must be
solved for the unknown functions S and $\hat{\rho}$.  Now a straightforward calculations shows
that, setting ${\bf (D70)}\,\,\psi(q,t)=\sqrt{\hat{\rho}(q,t)}exp](i/\hbar)S(q,t)]$, the quantity
$\psi$ obeys a linear PDE (corrected from \cite{sa})
\bq\label{5.23}
i\hbar\pp_t\psi=\frac{1}{2m}\left\{\left[\frac{i\hbar\pp_i\sqrt{g}}{\sqrt{g}}+A_i
\right]g^{ik}(i\hbar\pp_k+A_k)\right\}\psi+\left[V-\gag\frac{\hbar^2}{m}\dot{R}\right]\psi=0
\end{equation}
where only the Riemannian curvature $\dot{R}$ is present (any explicit reference to the gauge
vector $\phi_i$ having disappeared).  \eqref{5.23} is of course the SE in curvilinear coordinates
whose invariance under point transformations is well known.  Moreover {\bf (D70)} shows that
$|\psi|^2=\hat{\rho}(q,t)$ is the invariant probability density of finding the particle in the
volume element $d^nq$ at time t.  Then following Nelson's arguments that the SE together with
the density formula contains QM the present theory is physically equivalent to traditional
nonrelativistic QM.  One sees also from {\bf (D70)} and \eqref{5.23} that the time independent
SE is obtained via ${\bf (D71)}\,\,S=S_0(q)-Et$ with constant E and $\hat{\rho}(q)$.  In this
case the scalar curvature of space becomes time independent; since starting data at $t_0$ is 
meaningless one replaces {\bf (D65)} with a condition $\int_M\hat{\rho}(q)\sqrt{g}d^nq=1$.
\\[3mm]\indent
{\bf REMARK 4.1.}
We recall (cf. \cite{c7}) that in the nonrelativistic context the quantum potential has the form
$Q=-(\hbar^2/2m)(\pp^2\sqrt{\rho}/\sqrt{\rho})\,\,(\rho\sim\hat{\rho}$ here) and in more
dimensions this corresponds to $Q=-(\hbar^2/2m)(\gD\sqrt{\rho}/\sqrt{\rho})$.  In Section 5.2 we
have a SE involving $\psi=\sqrt{\rho}exp[(i/\hbar)S]$ with corresponding HJ equation \eqref{5.22}
which corresponds to the flat space 1-D ${\bf (D72)}\,\,S_t+(s')^2/2m+V+Q=0$ with continuity
equation $\pp_t\rho+\pp(\rho S'/m)=0$ (take $A_k=0$ here).  The continuity equation in {\bf (D69)}
corresponds to $\pp_t\rho+(1/m\sqrt{g})\pp_i[\rho\sqrt{g}g^{ik}(\pp_kS)]=0$.  For $A_k=0$
\eqref{5.22} becomes ${\bf (D73)}\,\,\pp_tS+(1/2m)g^{ik}\pp_iS\pp_kS+V-\gag(\hbar^2/m)R=0$.
This leads to an identification ${\bf (D74)}\,\,Q\sim-\gag(\hbar^2/m)R$ where R is the Ricci
scalar in the Weyl geometry (related to the Riemannian curvature built on standard Christoffel
symbols via {\bf (D58)}).  Here $\gag=(1/6)[(n-2)(n-2)]$ as in {\bf (D50)} which for $n=3$
becomes $\gag=1/12$; further by {\bf (D66)} the Weyl field $\phi_i=-\pp_i log(\rho)$.
Consequently (see also Remark 4.3).
\begin{proposition}
For the SE \eqref{5.23} in Weyl space the quantum potential is $Q=-(\hbar^2/12m)R$ where R
is the Weyl-Ricci scalar curvature.  For Riemannian flat space $\dot{R}=0$ this becomes
via {\bf (D67)}
\bq\label{5.24}
R=\frac{1}{2\gag\sqrt{\rho}}\pp_ig^{ik}\pp_k\sqrt{\rho}\sim\frac{1}{2\gag}
\frac{\gD\sqrt{\rho}}{\sqrt{\rho}}\Rightarrow
Q=-\frac{\hbar^2}{2m}\frac{\gD\sqrt{\rho}}{\sqrt{\rho}}
\end{equation}
as is should and the SE \eqref{5.23} reduces to the standard SE $i\hbar\pp_t\psi=
-(\hbar^2/2m)\gD\psi+V\psi$ ($A_k=0$).
$\hfill\bs$
\end{proposition}
\indent
{\bf REMARK 4.2.}
We recall next from \cite{c7} that
the Fisher information connection to the SE involves a classical ensemble
with particle mass m moving under a potential V
${\bf (D75)}\,\,
S_t+\frac{1}{2m}(S')^2+V=0;\,\,P_t+\frac{1}{m}\pp(PS')'=0$
where S is a momentum potential; note that no quantum potential is present but this will
be added on in the form of a term $(1/2m)\int dt(\gD N)^2$ in the Lagrangian which measures
the strength of fluctuations.  
This can then be specified in terms of the probability
density P leading to a SE (cf. \cite{gw,hl0,hl1,hl2,rg}).  One can also approach this via
(1-dimension for simplicity)
\bq\label{2.88}
S_t+\frac{1}{2m}(S')^2+V+\frac{\gl}{m}\left(\frac{(P')^2}{P^2}-\frac{2P''}{P}\right)=0
\end{equation}
Note that $Q=-(\hbar^2/2m)(R''/R)$ becomes for $R=P^{1/2}$ ${\bf (D76)}\,\,
Q=-(2\hbar^2/2m)[(2P''/P)-(P'/P)^2]$.  Thus the addition of the
Fisher information serves to quantize the classical system.
One also defines an information entropy (IE) via ${\bf (D77)}\,\,{\mf S}=-\int
\rho log(\rho)d^3x\,\,(\rho=|\psi|^2)$ leading to 
\bq\label{2.99}
\frac{\pp{\mf S}}{\pp t}=\int (1+log(\rho))\pp(v\rho)\sim \int \frac{(\rho')^2}{\rho}
\end{equation}
modulo constants involving $D\sim \hbar/2m$.  ${\mf S}$ is typically not conserved and 
$\pp_t\rho=-\na\cdot(v\rho)\,\,(u=D\na log(\rho)$ with $v=-u$ corresponds to standard
Brownian motion with
$d{\mf S}/dt\geq 0$.  Then high IE production corresponds to rapid flattening of the
probability density.  Note here also that ${\mf F}\sim -(2/D^2)\int \rho Qdx=\int
dx[(\rho')^2/\rho]$ is a functional form of Fisher information.
This leads one to conjecture that for the SE \eqref{5.23} in Weyl space there is a direct
connection between the Ricci-Weyl curvature and Fischer information in the form of the
quantum potential; this in turn suggests a connection between information entropy and
curvature.$\hfill\bs$
\\[3mm]\indent
{\bf REMARK 4.3.}
The formulation above from \cite{sa} was modified in \cite{sb} to a derivation of the
Klein-Gordon (KG) equation via an average action principle with the restrictions of Weyl
geometry released.  The spacetime geometry was then obtained from the action principle to
obtain Weyl connections with a gauge field $\phi_{\mu}$.  The Riemann scalar curvature 
$\dot{R}$ is then related to the Weyl scalar curvature R via an equation ${\bf (D78)}\,\,
R=\dot{R}-3[(1/2)g^{\mu\nu}\phi_{\mu}\phi_{\nu}+(1/\sqrt{-g})\pp_{\mu}(\sqrt{-g}g^{\mu\nu}
\phi_{\nu}]$.  Explicit reference to the underlying Weyl structure disappears in the
resulting SE (as in \eqref{5.23}).  The HJ equation in \cite{sb} has the form (for
$A_{\mu}=0$ and $V=0$) ${\bf (D79)}\,\,g^{\mu\nu}\pp_{\mu}S\pp_{\nu}S=m^2-(R/6)$ so in some
sense (recall here $\hbar=c=1$) ${\bf (D80)}\,\,m^2-(R/6)\sim {\mf M}^2$ (via \eqref{2.3})
where ${\mf M}^2=m^2exp(Q)$ and $Q=(\hbar^2/m^2c^2)(\bx\sqrt{\rho}/\sqrt{\rho})\sim
(\bx\sqrt{\rho}/m^2\sqrt{\rho})$ via \eqref{2.6} (for signature $(-,+,+,+)$).  Thus
for $exp(Q)\sim 1+Q$ one has ${\bf (D81)}\,\,m^2-(R/6)\sim m^2(1+Q)\Rightarrow (R/6)\sim
-Qm^2\sim -(\bx\sqrt{\rho}/\sqrt{\rho})$.  This agrees also with \cite{cas} where the whole
matter is analyzed incisively.
$\hfill\bs$

\end{document}